\documentclass[10pt,  letterpaper]{IEEEtran}
\usepackage{todonotes}
\usepackage[nocompress]{cite}
\usepackage{url}
\usepackage{enumitem}
\usepackage{epstopdf}
\PrependGraphicsExtensions{.tif, .tiff}
\usepackage{graphicx}
\usepackage{caption}
\usepackage{amsmath}
\usepackage{tabu}
\usepackage{multirow}
\usepackage{adjustbox}
\usepackage{mathptmx}
\usepackage{amsfonts}
\usepackage{algpseudocode}
\usepackage{soul}
\usepackage[bookmarks=false]{hyperref}
\usepackage{footnote}
\usepackage{lipsum}
\usepackage[ruled,vlined, linesnumbered]{algorithm2e}
\usepackage{subcaption}
\usepackage{listings}
\usepackage{tikz}
\usetikzlibrary{positioning}
\soulregister\cite7
\soulregister\ref7
\soulregister\pageref7
\soulregister\secref7
\soulregister\tabref7

\definecolor{ao}{rgb}{0.0, 0.5, 0.0}
\definecolor{db}{rgb}{0.2, 0.2, 0.6}
\definecolor{cadmiumgreen}{rgb}{0.0, 0.42, 0.24}

\newcommand{\tabref}[1]{Table~\ref{#1}}

\newcommand{\secref}[1]{Section~\ref{#1}}

\SetAlFnt{\footnotesize}

\begin{document}
\title{Architecture-Specific Performance Optimization of Compute-Intensive FaaS Functions}
\author{\IEEEauthorblockN{Mohak Chadha\IEEEauthorrefmark{1}, Anshul Jindal\IEEEauthorrefmark{1}, Michael Gerndt\IEEEauthorrefmark{1}\\}
\IEEEauthorblockA{\IEEEauthorrefmark{1}Chair of Computer Architecture and Parallel Systems, Technische Universit{\"a}t M{\"u}nchen \\
Garching (near Munich), Germany \\} 
Email: mohak.chadha@tum.de, jindal@in.tum.de, gerndt@in.tum.de}

\maketitle


\begin{abstract}

FaaS allows an application to be decomposed into functions that are executed on a FaaS platform. The FaaS platform is responsible for the resource provisioning of the functions. Recently, there is a growing trend towards the execution of compute-intensive FaaS functions that run for several seconds. However, due to the billing policies followed by commercial FaaS offerings, the execution of these functions can incur significantly higher costs. Moreover, due to the abstraction of underlying processor architectures on which the functions are executed, the optimization of these functions is challenging. As a result, most FaaS functions use pre-compiled libraries generic to x86-64 leading to performance degradation. In this paper, we examine the underlying processor architectures for Google Cloud Functions (GCF) and determine their prevalence across the 19 available GCF regions. We modify, adapt, and optimize a representative set of six compute-intensive FaaS workloads written in Python using Numba, a JIT compiler based on LLVM, and present results wrt performance, memory consumption, and costs on GCF. Results from our experiments show that the optimization of FaaS functions can improve performance by $18.2$x (geometric mean) and save costs by $76.8$\% on average for the six functions. Our results show that optimization of the FaaS functions for the specific architecture is very important. We achieved a maximum speedup of $1.79$x by tuning the function especially for the instruction set of the underlying processor architecture.



\end{abstract}


\IEEEpeerreviewmaketitle
\thispagestyle{empty}

\section{Introduction}
\label{sec:intro}
Since the introduction of AWS Lambda\cite{AWSLambda} by Amazon in 2014, serverless computing has grown to support a wide variety of applications such as machine learning~\cite{serverlessfl}, map/reduce-style jobs~\cite{jonas2017occupy}, and compute-intensive scientific workloads~\cite{chard2020funcx, scientific, kim2019functionbench, jindal2021function}. Function-as-a-Service (FaaS), a key enabler of serverless computing allows a traditional monolithic application to be decomposed into fine-grained functions that are executed in response to event triggers or HTTP requests~\cite{lynn2017preliminary} on a FaaS platform. Most commercial FaaS platforms such as AWS Lambda, Google Cloud Functions (GCF)~\cite{CloudFun36} enable the deployment of functions along with a list of static dependencies. The FaaS platform is responsible for generating containers using the static dependencies and the isolation, execution of these containers. These containers are commonly referred to as function instances.

\begin{figure}[t]
\centering
\includegraphics[width=\columnwidth]{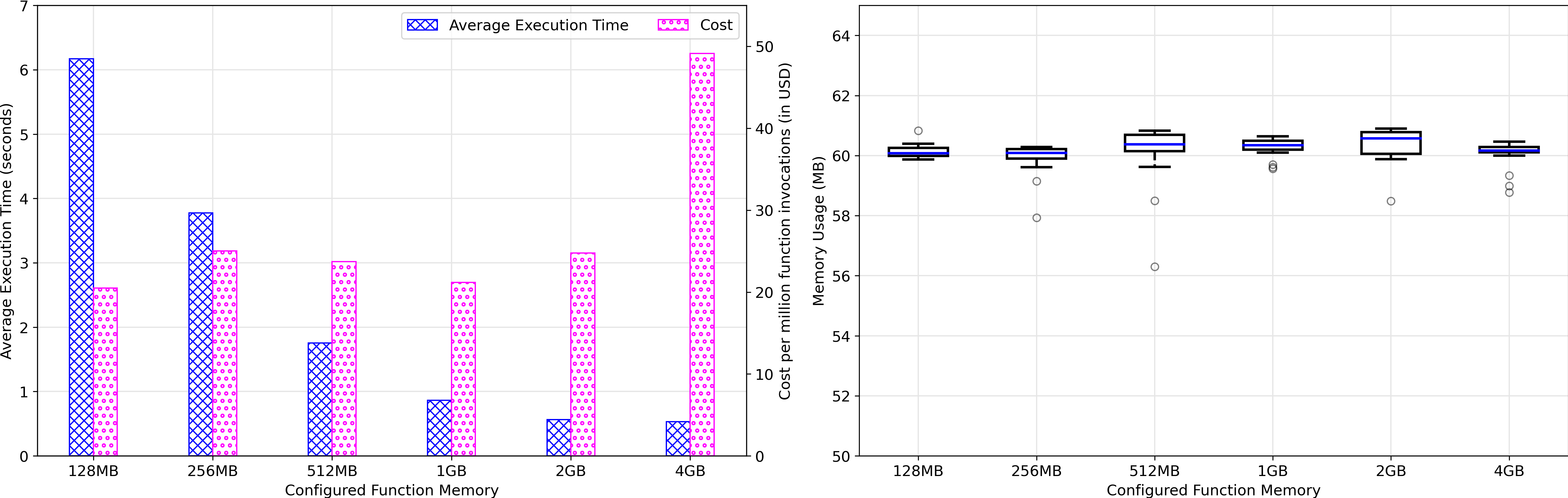}
     \caption{Average execution time, cost, and memory consumption for the \texttt{Floatbenchmark}~\cite{kim2019functionbench} when deployed with different memory configurations on GCF (\texttt{us-west2} region).}
\label{fig:cost_compute_eval}
\end{figure}

FaaS platforms follow a process-based model for resource management, i.e., each function instance has a fixed number of cores and quantity of memory associated with it~\cite{wang2018peeking}.  While today's commercial FaaS platforms such as Lambda, GCF abstract details about the backend infrastructure management away from the user, they still expose the application developers to explicit low-level decisions about the amount of memory  to  allocate  to  a  respective  function. These decisions affect the provisioning characteristics of a FaaS function in two ways. First, the amount of CPU provisioned for the function, i.e., some providers increase the amount of compute available to the function when more memory is assigned~\cite{daviducc, behind}. Selecting an appropriate memory configuration is an optimization problem due to the trade-offs between decreasing function execution time with increasing memory configuration and costs. Moreover, assigning more memory than desired can lead to significant resource over-provisioning and reduced malleability~\cite{spillner2020resource}.  Second, the addition of a per-invocation duration-utilization product fee measured in GB-Second (and GHz-Second with GCF~\cite{princinggcf}). 
FaaS is advertised as a pay-per-use model, where the users are billed based on the execution time of the functions measured typically in 100ms (GCF) or 1ms (Azure Functions~\cite{AzureFun16}, Lambda) intervals. As a result, for compute-intensive functions that require more than the minimum amount of memory the duration-utilisation component fee can lead to significantly higher costs. For instance, Figure~\ref{fig:cost_compute_eval} shows the comparison between the average execution time and cost~\cite{princinggcf} (excluding free tiers and networking) for the \texttt{Floatbenchmark}~\cite{kim2019functionbench} when deployed on GCF for the different available memory profiles. Although the average execution time decreases when more memory is configured, the cost increases. Moreover, the memory utilized per function instance is $60$MB as shown in Figure~\ref{fig:cost_compute_eval} leading to significant memory under-utilization. Improving the performance of compute-intensive FaaS applications can lead to reduction in execution time, memory over-provisioning, and thus reduced costs. 

While compute-intensive applications are written in a wide variety of high-level languages such as Java, R, and Julia. In this paper, we focus on Python since it is t widely used high-level programming languages for compute-intensive workloads such as image-processing, logistic regression, and scientific applications such as High Energy Physics Analysis~\cite{coffea}. Furthermore, it is supported by all major commercial FaaS platforms. To facilitate the performance improvement of applications written in Python several approaches exist. These include using an alternative Python interpreter such as PyPy~\cite{pypy}, Pyston~\cite{pyston}, and Pyjion~\cite{pyjion} or using a Python to C/C++ transpiler such as Cython~\cite{cython}, and Nuitka~\cite{Nuitika}. Using a replacement Python interpreter has the disadvantage that it has it's own ecosystem of packages which are significantly limited. Disadvantages of using a transpiler is that it offers limited static analysis, and that the code has to be compiled Ahead-of-Time (AOT). This leads to under-specialized and generic code for a particular CPU's architectural family (such as \texttt{x86-64}) or can cause code bloating to cover all possible variants~\cite{bloat}. To this end, we utilize Numba~\cite{numba}, a  Just-in-Time (JIT) compiler for Python based on LLVM~\cite{llvm} for optimizing and improving the performance of compute-intensive FaaS functions.  

On invocation of a deployed function, the function instances are launched on the FaaS platform's traditional Infrastructure as a Service (IaaS) virtual machines (VM) (microVMs~\cite{firecracker} in Lambda) offerings. However, the provisioning of such VMs is abstracted away from the user. As a result, the user is not aware of the details of the provisioned VMs such as the CPU architecture and the number of virtual CPUs (vCPUs). This makes optimizing FaaS applications challenging. 

Identification of the set of architectures dynamically used in current commercial FaaS platforms is  important for the performance optimization of FaaS functions. Previous works~\cite{wang2018peeking, behind} have reported the presence of Intel based processors ranging from Sandy Bridge-EP to Skylake-SP architectures in the provisioned VMs However, due to the rapid development in FaaS offerings of major cloud providers, and to offer updated insights, we investigate the current CPU processor architectures for GCF.

Our key contributions are:
\begin{itemize}
    \item We investigate the current CPU architectures present in GCF across the different regions.
    \item We analyze the impact of heterogeneity in the underlying processor architectures on the performance of a FaaS function.
    \item  We modify, adapt, and optimize a subset of six FaaS workloads\footnote{\url{https://github.com/kky-fury/Optimizing_FaaS_Workloads}} from FunctionBench~\cite{kim2019functionbench}, and the Python performance benchmark suite (Pyperf)~\cite{pyperf} using Numba. Although, the modified code is generic and can be used with any cloud provider, we use GCF in this work due to the availability of credits.
    \item We deploy the optimized workloads on GCF for the different memory profiles and analyze the impact on performance, costs, and memory consumption.
\end{itemize}

\thispagestyle{empty}

\begin{figure}[t]
\centering
\includegraphics[width=\columnwidth]{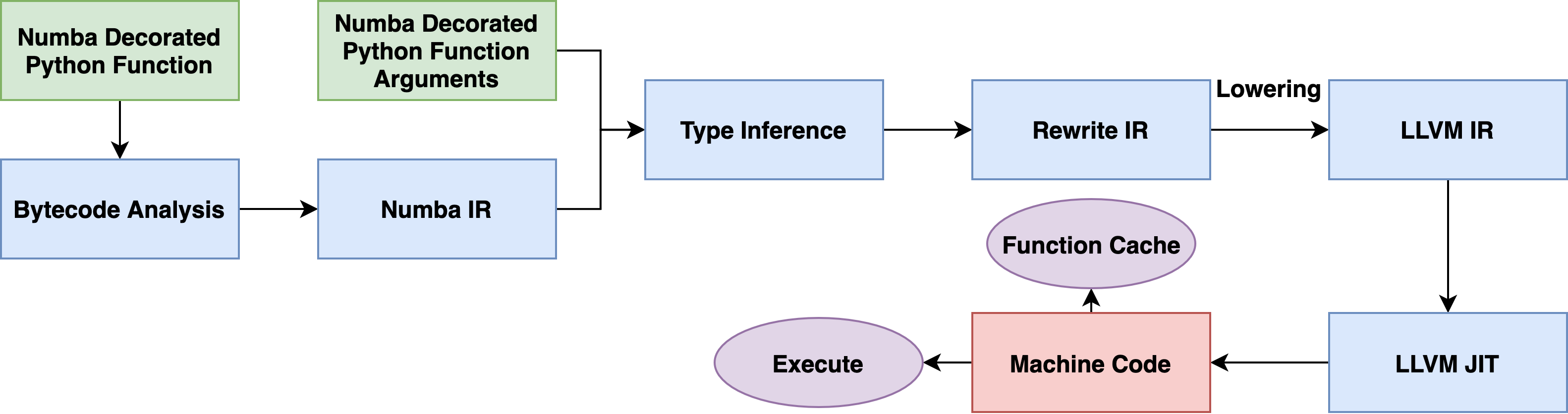}
\captionsetup{justification=centering}
\caption{Numba~\cite{numba} compilation workflow.}
\label{fig:workflow_numba}
\end{figure}

The rest of this paper is organized as follows. \S\ref{sec:background} gives a brief overview of Numba. In \S\ref{sec:relatedwork}, the current techniques for optimizing FaaS and previous works that investigated the backend infrastructure in major cloud provider's FaaS offerings are described. \S\ref{sec:methodology} describes our methodology for performance measurement, FaaS workloads used in this work, and our strategy for optimizing and maximizing the performance of the selected workloads with Numba. In \S\ref{sec:platformarch}, the different processor architectures we identified in the provisioned VMs across all GCF regions and the key differences in their microarchitectures that can impact the performance of functions optimized using Numba are described. In \S\ref{sec:results}, we present our evaluations results for the optimized FaaS workloads as compared to their native implementations in terms of performance, memory consumption, and costs. \S\ref{sec:conclusion} concludes the paper and presents an outlook. 

\section{Background}
\label{sec:background}

Numba~\cite{numba} is a function-at-a-time Just-in-Time (JIT) compiler for Python that is best suited for compute-intensive code that uses Numpy~\cite{harris2020array}, or scalar numerical code with loops. In contrast to Pypy~\cite{pypy}, Pyston~\cite{pyston}, and Pyjion~\cite{pyjion} it is implemented as a library and can be dynamically loaded by applications that use the native Python interpreter. To compile a native Python function to machine code using Numba, the user annotates the function using Python decorators (\texttt{jit, or njit}). The decorator replaces the function object with a special object that triggers compilation when the decorated function is called. 

Figure~\ref{fig:workflow_numba} shows the compilation workflow of a decorated function using Numba. In the first step, the function bytecode is analyzed. This includes recovering control flow information, disassembling the bytecode, and converting the native stack machine into a register machine (assigning virtual registers). Following this, the bytecode is translated into Numba IR which is a higher-level representation of the function logic than the native bytecode. To infer the types of the function arguments and variables, local type inference is applied on the generated Numba IR by building data dependency graphs. The function signatures are encoded and stored in a function registry. This is done to avoid recompilation of the decorated function if it is called again with different arguments of the same type. After type inference, several high-level optimizations such as deferring loop specializations and generation of array expressions are performed on the generated Numba IR. Following this, the rewritten Numba IR is translated (lowered) to LLVM IR. For converting the generated LLVM IR to machine code, Numba uses the high-quality compiler back-end with JIT support provided by LLVM~\cite{llvm}. Finally, the generated machine code is executed. To prevent recompilation and reduce overhead on future runs of the same function, Numba supports file-based caching of the generated machine code. This can be done by passing an argument to the Python decorator.

Note that, the generated machine code can be executed without the global interpreter lock (GIL) in Python, and thus can run parallel threads. In this paper, we utilize the Intel Thread Building Blocks~\cite{tbb} library, supported by Numba, to parallelize and optimize certain FaaS functions~\cite{anderson2017parallelizing}. Numba also provides support for generating code for accelerators such as Nvidia/AMD GPUs using NVVM~\cite{nvvm} and HLC~\cite{hsa}. Using GPUs for accelerating FaaS functions~\cite{gpuserverless} is our interest for the investigation in the future, but is out of scope for this work.




\section{Related Work}
\label{sec:relatedwork}
\textbf{FaaS Optimizations}. Majority of the previous works~\cite{agile, faasm, fuerst2021faascache} have focused on optimizing the cold start problem associated with FaaS. Mohan et al.~\cite{agile} identify the creation of network namespaces during container startup as the major reason for overhead for concurrent function invocations. Towards this, they propose the usage of Pause Containers (PCs), i.e., a set of pre-created containers with cached networking endpoints, thereby removing network creation from the critical path. Shillaker et al.~\cite{faasm} propose Faasm which uses the software fault isolation provided by WebAssembly to speed up the creation of a new execution environment. However, since it relies on language-level rather that container-based isolation, it makes it's integration and usage with public cloud providers challenging. Fuerst et al.~\cite{fuerst2021faascache} develop FaasCache, based on OpenWhisk, that implements a set of caching-based keep-alive policies for reducing the overhead due to function cold-starts. In contrast to previous works, we optimize the performance of a representative set of common FaaS workloads and present benefits/tradeoffs in terms of performance, memory consumption, and costs when deployed on a public cloud provider, i.e., GCF.


\begin{figure}[t]
\centering
\includegraphics[width=\columnwidth]{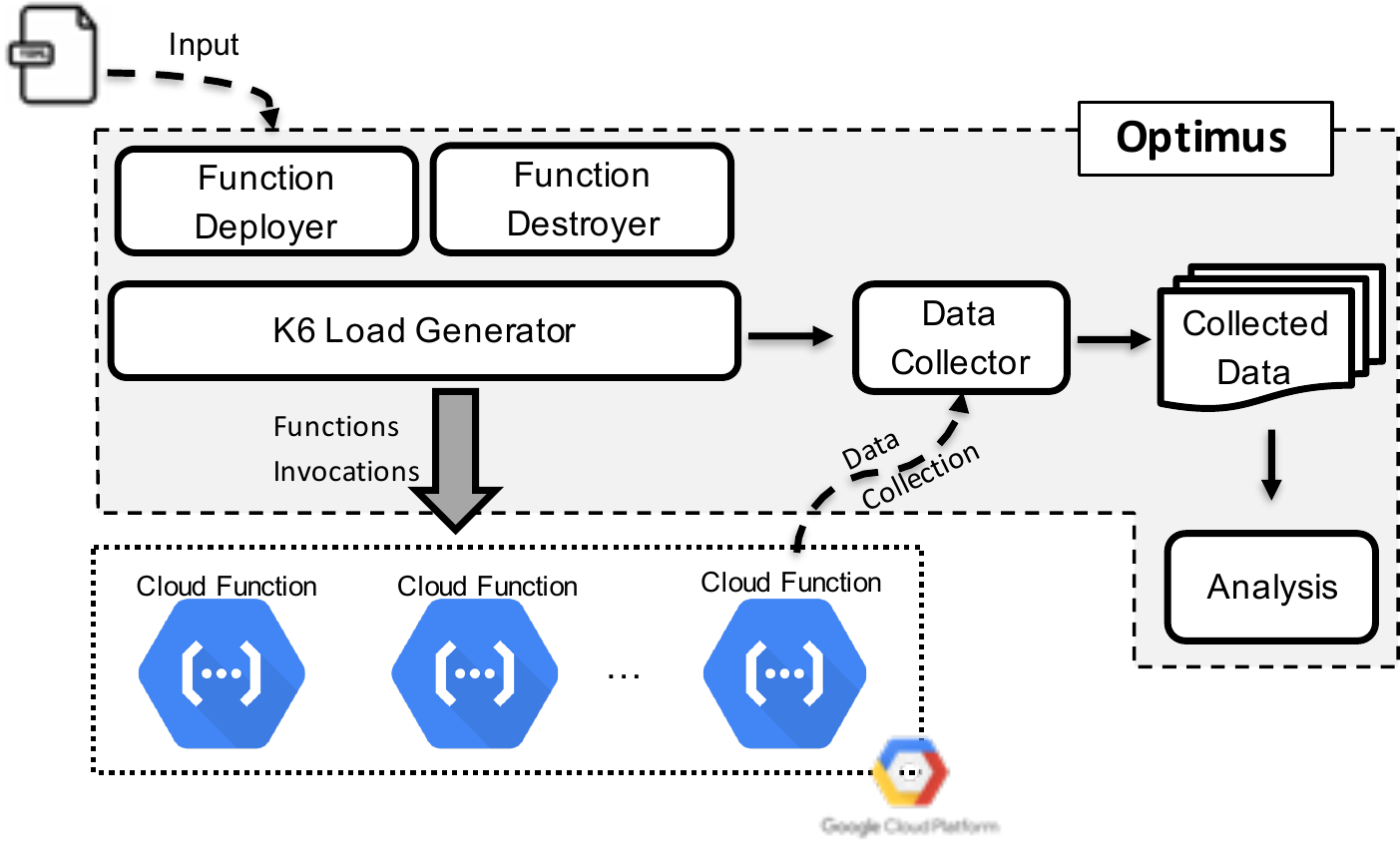}
\captionsetup{justification=centering}
\caption{Architecture of our benchmarking and data acquisition tool \textit{Optimus}.}
\label{fig:measuremtarch}
\end{figure}

\textbf{Understanding the Backend Infrastructure in Commercial FaaS Platforms}. The most notable works in this domain have been~\cite{wang2018peeking, behind}. Wang et al.~\cite{wang2018peeking} performed an in-depth study of resource management and performance isolation with three popular serverless computing providers: AWS Lambda, Azure Functions, and GCF. They
show that the provisioned VMs across the different platforms have great heterogeneity wrt the underlying processor architectures and configuration such as number of virtual CPUs. Kelly et al.~\cite{behind} provide an updated view on the VM toplogy of the major FaaS platforms including IBM Cloud Functions. Furthermore, they investigate the effect of interference on the cloud platforms due to the generated user load over a period of one month. While these previous works have inspired some of the methodology of the experiments used in this work, there are some key differences. First, we identify the prevalence of different processor architectures in the provisioned VMs across the 19 different available GCF regions. Second, we demonstrate how the underlying VM configuration such as the number of vCPUs can be used for optimizing the performance of functions. Third, we demonstrate the effect of microarchitectural differences in the underlying processor architectures on the performance of FaaS functions.

\thispagestyle{empty}


\textbf{JIT Compilers for Native Python}. Besides Numba, there exist other JIT compilers such as Psyco~\cite{psyco}, and Unladen Swallow~\cite{swallow}. Psyco has a built-in compiler for the native Python interpreter and features it's own \texttt{x86}-only code generator. Swallow was sponsored by Google and aimed to modify and integrate the native Python interpreter with a JIT compiler based on LLVM. However, both of these projects have been discontinued. As a result, we use Numba in this work.

\section{Methodology and Benchmarks}
\label{sec:methodology}
In this section, we describe \textit{Optimus}, a Python-based tool for benchmarking and collecting metric data from functions deployed on GCF. Following this, we describe the FaaS workloads we used and optimized in this work. Finally, we describe our approach for optimizing and maximizing the performance of the selected workloads using Numba.

\subsection{Benchmarking and data acquisition}
\label{sec:dataacq}
To facilitate the deployment, deletion, benchmarking, and metric data acquisition of functions on GCF, we have developed \textit{Optimus}. It's architecture and different components are shown in Figure~\ref{fig:measuremtarch}. 
\textit{Optimus} takes a YAML file as input that specifies the GCF function configuration parameters (deployment region, memory configuration, maximum number of function instances, timeout etc.) for the function deployment, the function to be deployed, and configuration parameters for the load generator. Following this, the \textit{Function Deployer} which encapsulates the functionality of the \texttt{gcloud function} command-line tool deploys the function according to the specified parameters.

\begin{table}[t]
\caption{Collected GCF monitoring metrics. The metric data is sampled every 10 seconds.}
	\centering

    \begin{tabu}{|c|c|}
	\tabucline{-}

    \textbf{Metric} & \textbf{Description} \\\tabucline{-}
    Active instances & The number of active function instances. \\ \tabucline{-}
    Function Invocations & The number of function invocations. \\ \tabucline{-}
    Allocated Memory & Configured function memory \\ \tabucline{-}
    Execution time & The mean execution time of the function \\ \tabucline{-}
    Memory usage & The mean memory usage of the function. \\ \tabucline{-}
    
\end{tabu}

\label{tab:metrics}
\end{table}

For all our tests, we deploy a virtual machine (VM) to use \textit{Optimus} on a private Compute Cloud available in our Institute. The VM is configured with 10 vCPUs (Intel Skylake-SP) and $45$GB of RAM. To invoke and evaluate the performance of the deployed function, we use \texttt{k6}~\cite{perftesting}. \texttt{k6} is a developer-centric open-source load and performance regression testing tool. It uses a script for executing the test where the deployed function's HTTP(s) endpoint along with the request parameters are specified. As part of each \texttt{k6} test, two additional parameters are configured, i.e., Virtual Users (VUs), and duration. VUs are the entities in \texttt{k6} that execute the test and make HTTP(s) or WebSocket requests. Duration specifies the total time a test will run. The number of requests per second (RPS) generated by \texttt{k6} depends on the number of VUs and the time taken by each request to complete. The number of VUs and the duration of the test can be specified in the input YAML file.


To collect the metric data on completion of a function load test, we implement a monitoring client using the Google Cloud client library~\cite{gcfmonitoring}. The different monitoring metrics extracted as part of each test are shown in Table~\ref{tab:metrics}. Note that, the sampling rate for each metric is 10 seconds which is the granularity supported by GCF~\cite{gcfmonitoringlimit}. The collected metric data is written to a csv file by the monitoring client and stored in deployed VM's local storage. After the metric data is collected, the \textit{Function Destroyer} deletes the deployed function to free up the resources. The data collected from several functions is later collated and analyzed.

\subsection{FaaS workloads}
\label{sec:benchmarks}

\begin{table}[t]
\caption{FaaS workloads used and optimized.}
	\centering
\begin{adjustbox}{width=9cm,center}
    \begin{tabu}{|c|c|c|}
	\tabucline{-}

    \textbf{Category} & \textbf{Name}   & \textbf{Suite}\\\tabucline{-}
    Micro-benchmark & Floatbenchmark & FunctionBench~\cite{kim2019functionbench} \\\tabucline{-}
    Application & Montecarlo, Image processing  & PyPerf~\cite{pyperf},  FunctionBench~\cite{kim2019functionbench} \\\tabucline{-}
    ML model training & Logistic regression  & FunctionBench~\cite{kim2019functionbench} \\\tabucline{-}
    Scientific simulation & Nbody & PyPerf~\cite{pyperf} \\\tabucline{-}
    Data Modelling & Kerneldensityestimate (KDE) & Other \\\tabucline{-}

\end{tabu}
\end{adjustbox}
\label{tab:benchmarks}
\end{table}

To demonstrate the advantages of optimizing compute-intensive FaaS functions, we use a wide-variety of  workloads from different categories, i.e., Micro benchmark, application, ML model training, scientific simulation, and data modelling. The individual workloads and the suites to which they belong are shown in Table~\ref{tab:benchmarks}.

The \textit{Floatbenchmark} performs a series of floating point arithmetic operations, i.e, squareroot, sin, and, cos followed by a reduction operation on the calculated values. It takes a JSON file as input specifying the number of iterations and returns the aggregated sum. The native implementation uses the \texttt{math} Python module. The \textit{Image processing} application uses the Python \texttt{Pillow}~\cite{pillow} library to blur a RGB image using the Gaussian Kernel and then converts the blurred image to grayscale. Following this, the Sobel operator is applied to the grayscale image for edge detection. As input, the workload takes a JSON file specifying the URLs to the images. After completion of the function the modified images are written to a block storage. Montecarlo simulations are commonly used in various domains such as finance, engineering, and supply chain. It is a technique commonly used to understand the impact of risk and uncertainty in prediction and forecasting models. The function calculates the area of a disk by assigning multiple random values to two variables to generate multiple results and then averages the results to obtain an estimate. It takes a JSON file as input specifying the number of iterations for the computation and returns the estimated area.

Logistic regression is a popular linear statistical and machine learning technique commonly used for classification tasks. It uses a logistic function to model the probabilities describing the possible outcomes of a trial. The workload uses a  Numpy~\cite{harris2020array} implementation of the logistic regression algorithm to build classifiers for the Iris~\cite{iris} and Digits datasets~\cite{digit}. The NBody problem commonly used in astrophysics involves predicting the motion of celestial objects interacting with each other under the influence of gravity. It involves the evaluation of all pairwise interactions between the involved bodies. The workload simulates the interactions between five bodies, i.e., the Sun, Jupiter, Saturn, Uranus, and Neptune. It takes a JSON file as input, specifying the number of iterations for the simulation, initial positions of the bodies according to a predefined coordinate system and returns the positions of the bodies after the simulation.

Kernel density estimation is a statistical technique that is used to estimate the probability density function of the underlying distribution. It allows the creation of a smooth curve on the given dataset which can be used for the generation of new data. The workload uses the gaussian kernel to estimate the density function. The native implementation is written using Numpy. As input, it takes a JSON file specifying the size of the distribution, bandwidth (smoothing parameter) of the kernel, and evaluation point for computing the estimate. On completion, it returns the calculated estimate at the evaluation point.


\thispagestyle{empty}

\subsection{Optimizing and maximizing performance with Numba}
\label{sec:perf_tips}

Our strategies for optimizing the different FaaS workloads varied with each function. For instance, with the \textit{Floatbenchmark} it was sufficient to decorate the function with the Numba \texttt{@njit} decorator (\S\ref{sec:background}) to get optimal performance, while for other workloads we identified performance bottlenecks using the \texttt{line\_profiler} and implemented optimized kernels, i.e., we refactored the native implementation of the workloads to enable automatic optimization by Numba. Towards this, we made use of different decorators supported by Numba such as \texttt{@stencil} and additional libraries such as Intel Short Vector Math Library (SVML)~\cite{svml}, and Intel TBB~\cite{tbb}. The \texttt{@stencil} decorator allows the user to specify a fixed computational pattern according to which the array elements of an input array are updated. Numba uses the decorator to generate looping code for applying the stencil to the input array. We used this decorator in the \textit{Image processing} workload (\S\ref{sec:benchmarks}) for blurring the input image with the Gaussian Kernel.

An important aspect of optimizing compute-intensive functions is vectorization of loops to generate Single Instruction Multiple Data (SIMD) instructions. The LLVM backend in Numba offers auto-vectorization of loops as a compiler optimization pass. On successful vectorization, the compiler will generate SIMD instructions depending on underlying processor's supported SIMD instruction set such as Advanced Vector Extensions (AVX)-2, AVX-512 (\S\ref{sec:perf_aspects}). However, auto-vectorization can often fail if the code analysis detects code properties that inhibit SIMD vectorization (such as data dependencies within the loop) or if compiler heuristics (such as vectorization efficiency) determine that SIMD execution is not beneficial. To identify if our implemented code was vectorized and to investigate the reasons for non-vectorization, we analyzed the generated optimization report by LLVM. We found that the most common reason for non-vectorization of loops to be the division of two numbers. This is because according to the Python convention which is followed by Numba, a division of two numbers expands into a branch statement which raises an exception if the denominator is zero. Since the autovectorizer offered by LLVM always fails if branches are present inside the loop the code is not vectorized. We were able to ensure vectorization of such loops by adding \texttt{error\_model='numpy'} to the \texttt{@njit} decorator in Numba through which division by zero results in \texttt{NaN}. As a sanity check, we also checked the generated assembly code for the \texttt{@njit} decorated Python function through the \texttt{inspect\_asm()} functionality offered by Numba. To further enhance performance, we utilized the SVML library through the \texttt{icc\_rt} Python package. The SVML library provides SIMD intrinsics, i.e., functions that correspond to a sequence of one or more assembly instructions, for packed vector scalar math operations. On inclusion of the \texttt{icc\_rt} package, Numba configures the LLVM backend to use the offered intrinsic functions whereever possible. 

In this paper, we use the Intel TBB library (\S\ref{sec:background}) as a threading backend supported by Numba to parallelize the \textit{Floatbenchmark}, \textit{Montecarlo}, and individual kernels (gaussian blur, and RGB to gray conversion) of the \textit{Image processing} workload. This was done by adding \texttt{parallel=True} argument to the \texttt{@njit} decorator. On successful parallelization, Numba generates machine code that can run on multiple native threads. The other benchmarks were not parallelized due to data and loop-carried dependencies in the implemented kernels. We use the \texttt{tbb}\footnote{\texttt{version==2020.0.133}} Python package for TBB support.

\begin{table}[t]
\caption{Data collected from the \texttt{proc} filesystem of the provisioned VM on GCF.}
	\centering
    \begin{tabu}{|c|c|}
	\tabucline{-}

    \textbf{Attribute} & \textbf{System Information} \\\tabucline{-}
    vCPUs & Number of virtual CPUs configured in the VM. \\ \tabucline{-}
    CPU Model & CPU model present in the VM. \\ \tabucline{-}
    CPU Family & Family of processors to which the CPU belongs. \\ \tabucline{-}
    Total Memory & Total memory configured in the VM. \\ \tabucline{-}
\end{tabu}
\label{tab:attributes}
\end{table}






For most workloads, we also added the argument \texttt{fastmath=True} to the \texttt{@njit} decorator. This relaxes the IEEE 754 compliance for floating point arithmetic to gain additional perfomance. Furthermore, it permits reassociation of floating point operations which allows vectorization. Note that, for all workloads we assume double precision floating point operations and ensure that the resultant output from the native and the optimized code is same within a tolerance value. We open-source the code for the optimized FaaS workloads.


\thispagestyle{empty}



\section{Platform Architecture}
\label{sec:platformarch}
In this section, we describe our methodology for identifying the underlying processor architectures in GCF. Following this, we describe the key differences in the microarchitecture of the determined processors that can impact the performance of compute-intensive functions when optimized using Numba.

\subsection{Identifying processor architectures on provisioned VMs in GCF}
\label{sec:processor_arch}

The GCF service is regional, i.e., the infrastructure on which the function instance is launched varies across the different available regions~\cite{gcflocations}. Moreover, the billing also varies depending on where the function is deployed, i.e., Tier 1, and Tier 2 pricing~\cite{princinggcf}. Functions deployed on Tier 2 regions, e.g, \texttt{us-west2} have a higher duration-utilization product fee measured in GB-Second and GHz-Second as compared to functions deployed in Tier 1 regions. To investigate the different underlying processor architectures of the provisioned VMs across the 19 available GCF regions, similar to~\cite{wang2018peeking, behind}, we used the \texttt{proc} filesystem on Linux. Table~\ref{tab:attributes} shows the different attributes we read from the Linux \texttt{procfs}. We obtained the number of virtual CPUs present in the provisioned VM by counting the number of processors present in the \texttt{/proc/cpuinfo} file. The CPU model and family were obtained through specific fields present in the \texttt{/proc/cpuinfo} file. We obtained the total memory configured in the VM using the \texttt{MemTotal} attribute in the \texttt{/proc/meminfo} file. 

\begin{figure}[t]
\centering
\includegraphics[width=\columnwidth]{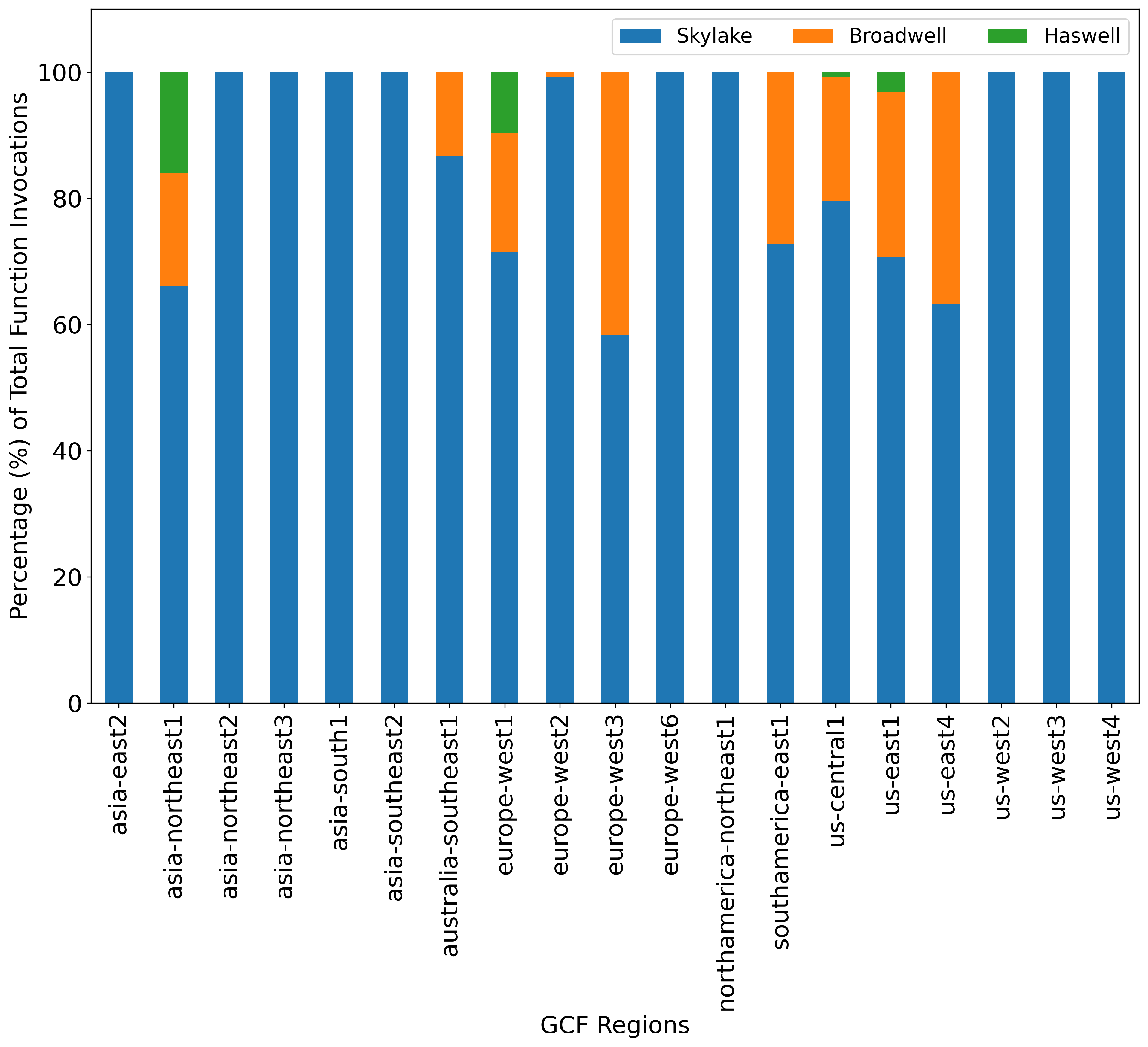}
\caption{The different Intel processor architectures across the 19 available GCF regions along with percentage of functions invoked on them.}
\label{fig:diffarchs}
\end{figure}

We implemented a function that reads the described attributes and collates them into a JSON response. Following this, we deployed the function for the different supported memory profiles at the time of the experiments\footnote{The experiments were performed in Feb-March 2021.}, i.e., $<128, 256, 512, 1024, 2048, 4096>$ MB across all the available regions using the function deployer component in \textit{Optimus} (\S\ref{sec:dataacq}). We fixed the number of virtual users and the duration of the test in \texttt{k6} to 60 and 1 minute respectively. As a result, multiple function instances were launched simultaneously to handle the requests. The obtained JSON reponses are stored on the deployed VM
as described in \S\ref{sec:dataacq}. We repeated the \texttt{k6} load test every two hours and collected the measurements for a period of two weeks, leading to more than a billion function invocations.

From the collected data, we found that across all regions the VMs provisioned were based on Intel Xeon CPUs. Although Google uses a proprietary hypervisor for running the function instances which hides the model name attribute from the Linux \texttt{procfs}, we were able to infer the different processor architectures using the model and family attributes~\cite{intelcpus}. Particularly, we found three different models from the same family \texttt{6}, i.e., \texttt{85-}Skylake, \texttt{79-}Broadwell, and \texttt{63-}Haswell. The family \texttt{6} represents Intel's Server CPU offerings and the numbers \texttt{85,79,63} are the different model numbers. Note that, the Intel processor architectures Cooper Lake and Cascade Lake also have the same model \texttt{85} as Skylake and belong to the same family. Due to the information abstracted by the Google's hypervisor it was not possible to distinguish between the different architectures. As a result, we classify it as Skylake. Similarly, it was not possible to uniquely identify the individual VMs as previously described by~\cite{wang2018peeking}, ~\cite{behind}.


In contrast to the results reported by~\cite{wang2018peeking, behind}, we did not find the architectures \texttt{(62,6)-}IvyBridge, \texttt{(45,6)-}SandyBridge on any of the provisioned VMs across all GCF regions. We believe since these models were launched in 2013~\cite{ivybridge} and 2012~\cite{sandybridge} respectively, they have been phased out. Figure~\ref{fig:diffarchs} shows the prevalence of the different architectures we found across the 19 available GCF regions. For a particular region, we combined the results for all the memory profiles. We found that Intel Skylake was the most prevalent architecture across all regions. Only for the regions \texttt{asia-northeast1, europe-west1, us-central1}, and \texttt{us-east1} we found function instances being launched on VMs with all the three processor architectures. We found the greatest heterogeneity in the \texttt{asia-northeast1} region with $16.1$\%, $17.9$\%, and $66$\% of the functions in that region being invoked on VMs with Haswell, Broadwell, and Skylake architectures respectively. For all regions, we found that irrespective of the configured memory profile the VMs were configured with 2GB of memory and 2 vCPUs. This was also true for a function configured with $4$GB of memory. As a sanity check, we wrote a simple function which allocates 3GB of memory when the function is configured with 4GB~\cite{awsblog}. This results in a heap allocation error. We believe that this is a bug and have reported it to Google.




\subsection{Key Microarchitectural Differences}
\label{sec:perf_aspects}
As described in \S\ref{sec:perf_tips}, a key aspect in performance optimization of compute-intesive applications on modern CPUs is the generation of SIMD instructions. While the Intel Skylake processor has several new microarchitectural features, which increase performance, scalability, and efficiency as compared to the Broadwell and Haswell architectures~\cite{schone2019energy}, in this paper, we focus only on differences in the SIMD instruction set.

The Intel Skylake processor supports the AVX-512 SIMD instruction set as compared to AVX-2 in both Broadwell and Haswell architectures. This means that each SIMD unit in Skylake has a width of 512 bits as compared to 256 bits in Broadwell and Haswell. As a result, with AVX-512 eight double precision or 16 single precision floating numbers can be used as input for vector instructions as compared to four and eight in Broadwell and Haswell respectively. Thus, doubling the number of FLOPS/cycle and improving performance. Note that, both AVX-2 and AVX-512 also support other datatypes such as long, short integers. 


On successful autovectorization the LLVM backend compiler used in Numba will try to generate SIMD instructions based on the highest available instruction set (\S\ref{sec:perf_tips}). The SIMD instruction set used can be easily identified by examining the assembly code of the  compiled jitted Numba function (\texttt{inspect\_asm()}). All AVX-512 instructions will use the \texttt{zMM} registers, while AVX-2 instructions will use the \texttt{yMM} registers. Note that, even though we classify the Intel Cascade and Cooper Lake processors (if present on GCF) as Skylake (\S\ref{sec:processor_arch}), the highest SIMD instruction set supported by them is AVX-512.


\thispagestyle{empty}


\section{Experimental Results}
\label{sec:results}
In this section, we evaluate the performance of the optimized FaaS workloads\footnote{We use the term workload and function interchangeably.} as compared to their native implementations and present results wrt average execution time, memory consumption, and costs. Following this, we investigate how the underlying heterogeneous processor architectures (\S\ref{sec:processor_arch}) effect the performance of a FaaS function. Furthermore, we demonstrate the importance of optimizing a FaaS function according to the SIMD instruction set of the underlying processor architecture.


\begin{table}[t]
\caption{Input configuration parameters for the individual FaaS workloads.}
	\centering
\begin{adjustbox}{width=6cm,center}
    \begin{tabu}{|c|c|}
	\tabucline{-}

    \textbf{Benchmark} & \textbf{Input configuration} \\\tabucline{-}
    Floatbenchmark & \texttt{100000} iterations. \\ \tabucline{-}
    Montecarlo & Forty million iterations. \\ \tabucline{-}
    Image processing & $4$ RGB images. \\ \tabucline{-}
    Logistic Regression & Iris, digits dataset. \\ \tabucline{-}
    Nbody & Fifty iterations. \\ \tabucline{-}
    KDE & Five million distribution size. \\ \tabucline{-}
  
\end{tabu}
\end{adjustbox}
\label{tab:config}
\end{table}

\begin{figure*}[t]
\centering
    \begin{subfigure}{0.49\textwidth}
    \centering
        \includegraphics[width=\columnwidth]{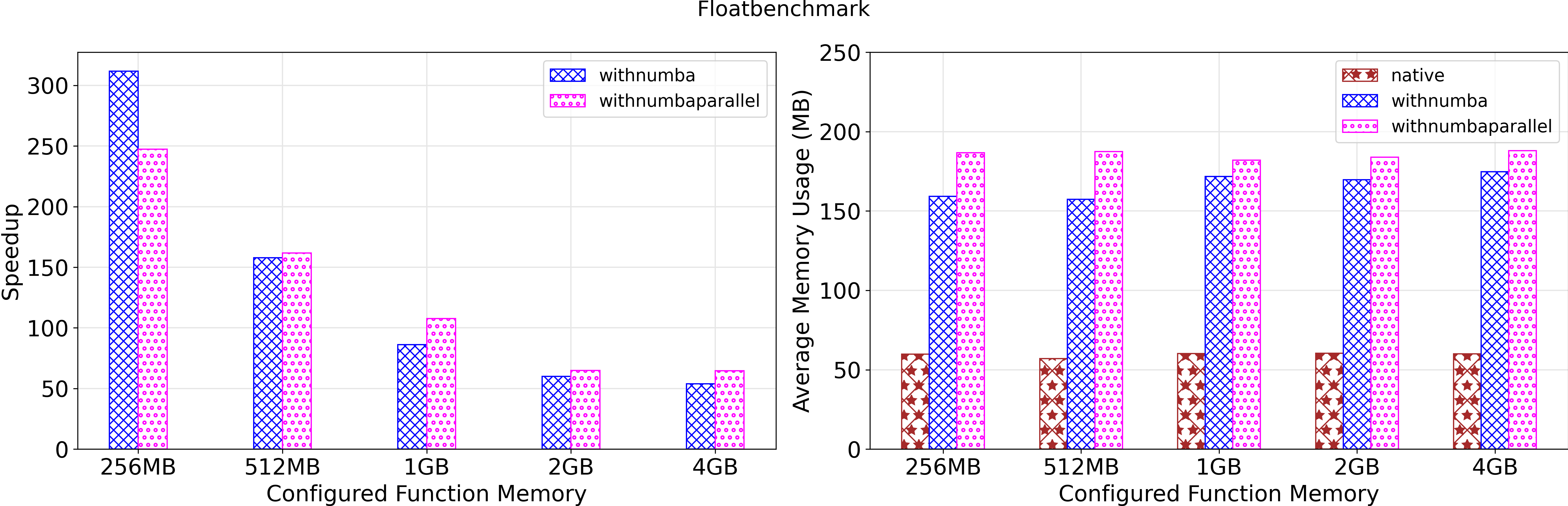}
        \caption{Floatbenchmark.}
        \label{perf_float}
    \end{subfigure}
    \begin{subfigure}{0.49\textwidth}
    \centering
        \includegraphics[width=\columnwidth]{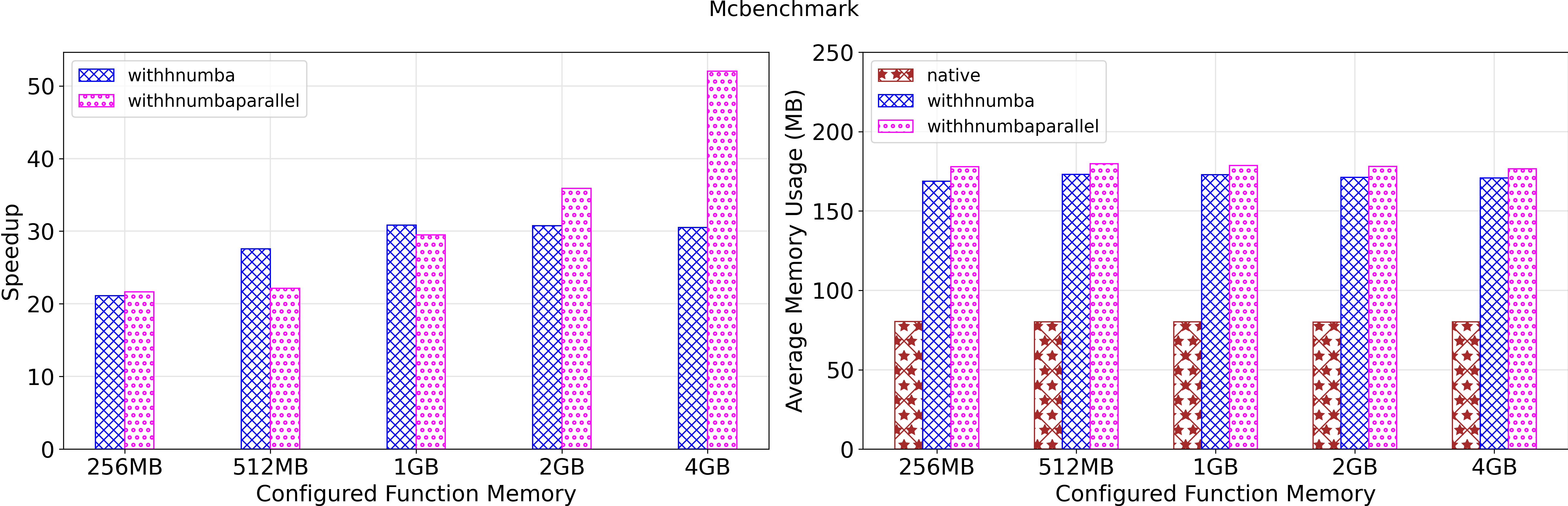}
        \caption{Mcbenchmark.}
        \label{perf_mcb}    
    \end{subfigure}
    \begin{subfigure}{0.49\textwidth}
    \centering
        \includegraphics[width=\columnwidth]{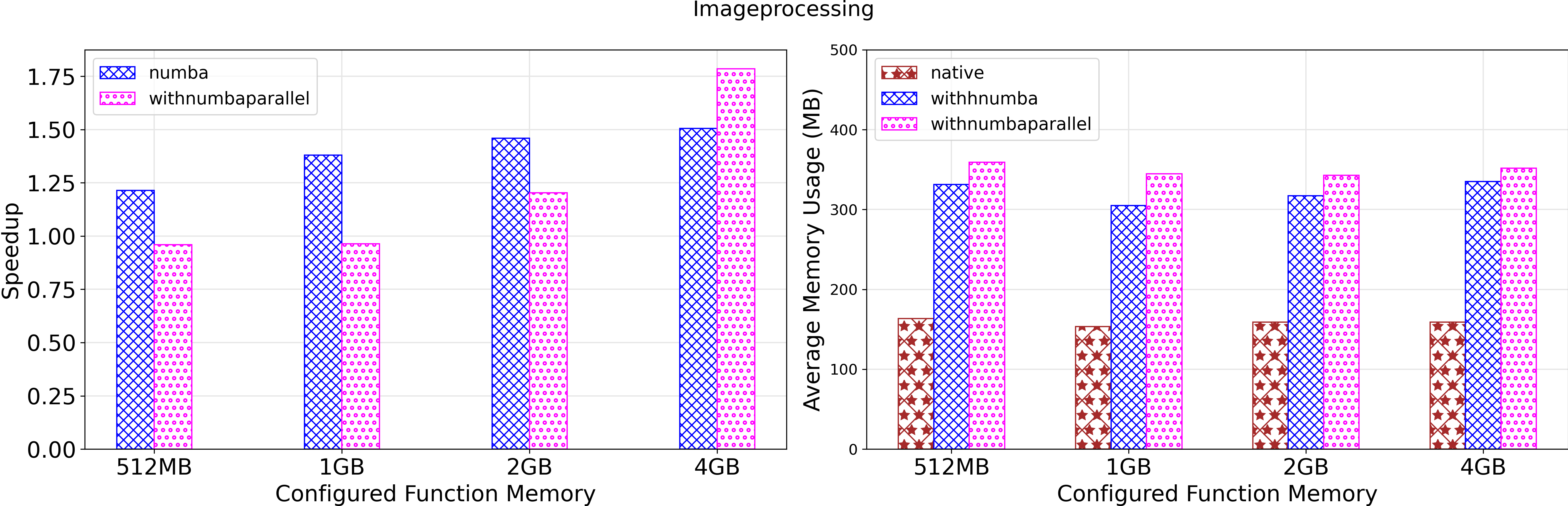}
        \caption{Imageprocessing}
        \label{perf_imgp}    
    \end{subfigure}
    \begin{subfigure}{0.49\textwidth}
    \centering
        \includegraphics[width=\columnwidth]{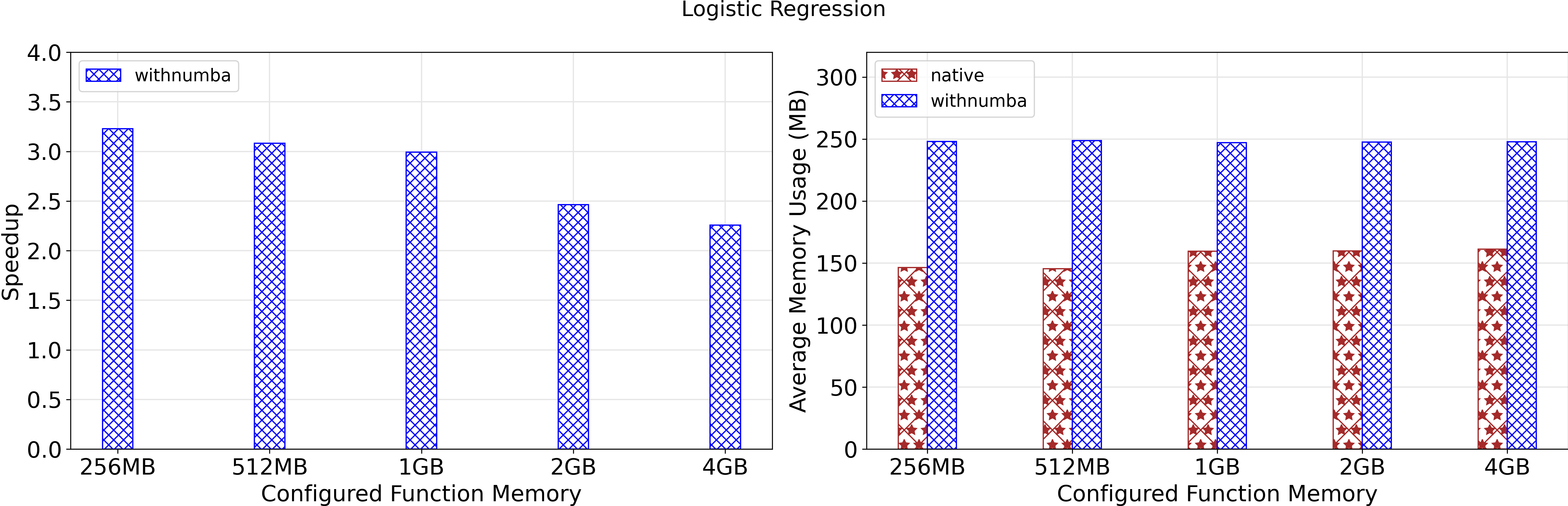}
        \caption{Logistic Regression}
        \label{perf_lr}    
    \end{subfigure}
    \begin{subfigure}{0.49\textwidth}
    \centering
        \includegraphics[width=\columnwidth]{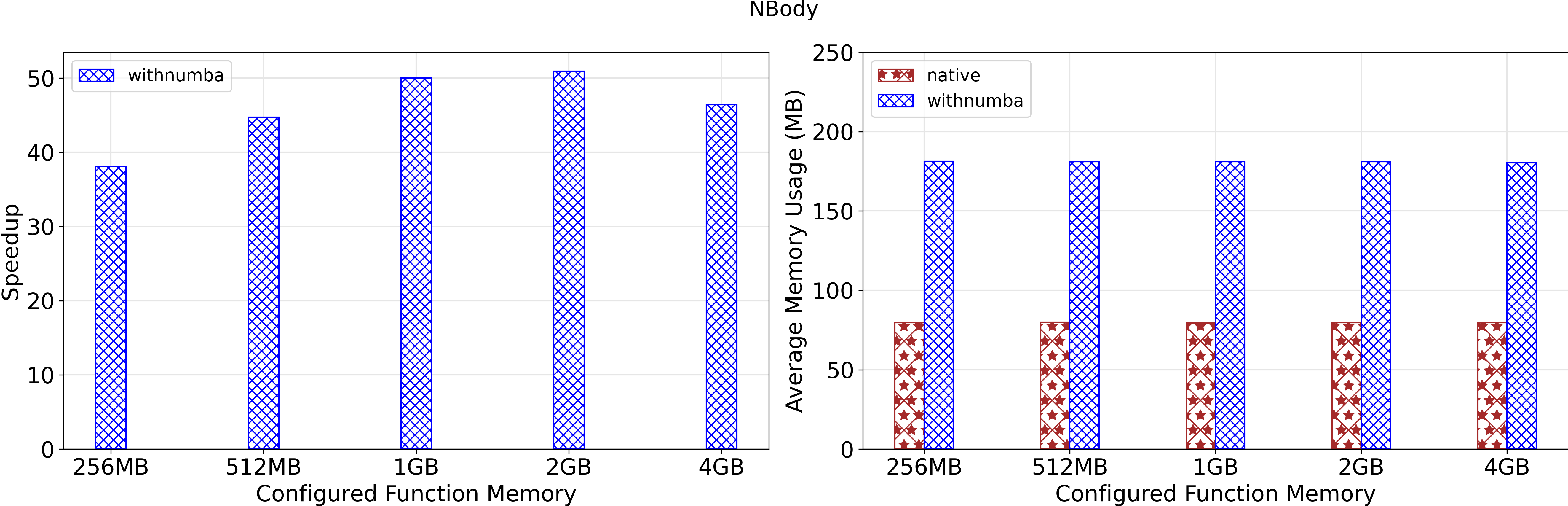}
        \caption{Nbody}
        \label{perf_nb}    
    \end{subfigure}
    \begin{subfigure}{0.49\textwidth}
    \centering
        \includegraphics[width=\columnwidth]{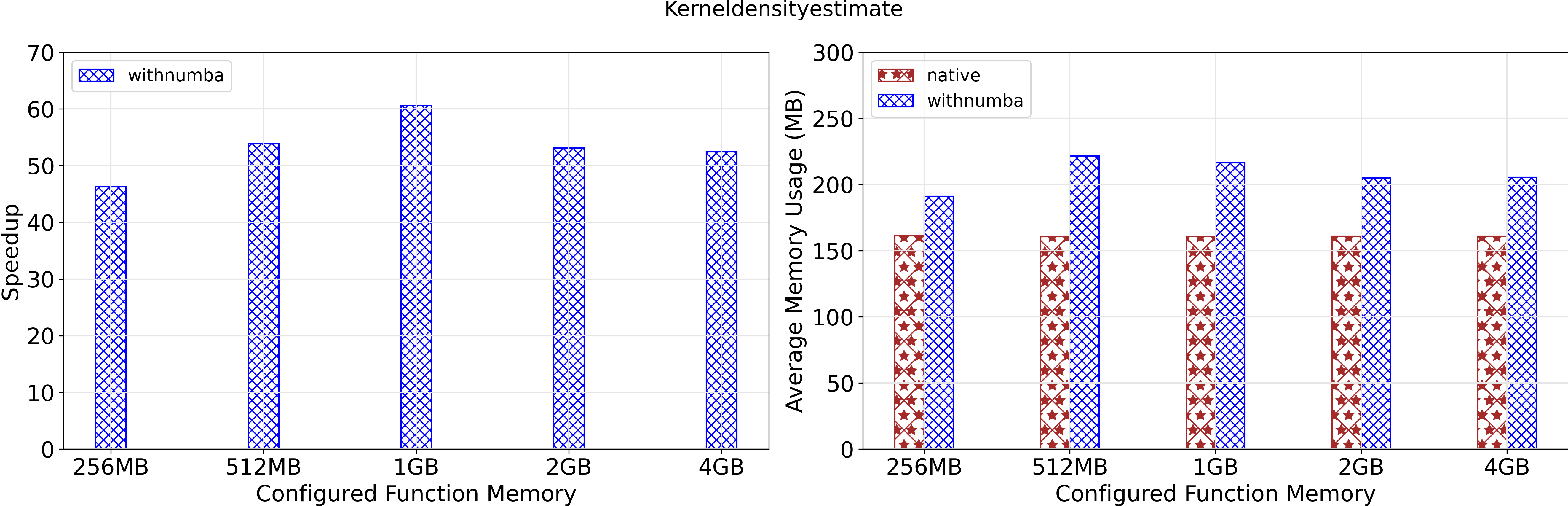}
        \caption{Kerneldensityestimate}
        \label{perf_kde}    
    \end{subfigure}
    \caption{The obtained speedup and average memory consumption of the six optimized FaaS workloads as compared to their native implementations for the different memory configurations on GCF. All functions were deployed on the \texttt{us-west2} region.}
    \label{fig:perfmemcomp}
\end{figure*}

\subsection{Experimental Configuration}
\label{sec:setup}
To compare the optimized and the native FaaS workloads wrt performance, memory consumption, and costs we deploy both versions 
on the \texttt{us-west2} GCF region for all the available memory profiles using \textit{Optimus} as described in \S\ref{sec:dataacq}. For all workloads, we set the maximum number of function instances to $50$ and the timeout to $300$ seconds. We chose \texttt{us-west2} since it was one of the regions where we observed homogeneous processor architecture, i.e., Skylake in the provisioned VMs (\S\ref{sec:processor_arch}). As configuration parameters to \texttt{k6}, we set the maximum number of VUs to $50$ and total duration of the load test to five minutes. For all our experiments, we repeated the \texttt{k6} test five times every two hours and then averaged the results. The individual input configuration parameters for each workload are shown in Table~\ref{tab:config}.

For all the optimized FaaS workloads, we enabled file-based caching of the compiled function machine code by adding the \texttt{cache=True} argument to the \texttt{@njit} decorator (\S\ref{sec:background})). We modified the Numba configuration to save the cached code in \texttt{/tmp} filesystem available for GCF. This was done to ensure that function instances provisioned on the same VM have access to the compiled machine code to avoid overhead due to recompilation. This behaviour was first reported by~\cite{behind}, where functions executing on the same VM could read a unique id written to a file in the \texttt{tmp} filesystem. From our experiments, we observed that caching improved the speedup by $1.2$x on average as compared to the non-cached version. The speedup was not much more significant because Numba jitted functions are stored in memory and retain their state between warm invocations. This means that recompilation of a Numba jitted function (with same function argument types) only occurs with a function cold start, i.e., when the execution environment is run for the first time. Moreover, for the parallelized FaaS functions, i.e., \textit{Floatbenchmark, Montecarlo}, and some kernels of the \textit{Image Processing} workload (\S\ref{sec:perf_tips}), we configured the number of TBB threads to two due to the availability of two vCPUs (\S\ref{sec:processor_arch}).

\thispagestyle{empty}

\subsection{Comparing performance and memory consumption}
\label{sec:perfmem}
For comparing the performance of the optimized FaaS workloads with their native implementations, we calculate the metric speedup. This is done by dividing the obtained average execution time of the native implementation by the obtained average execution time of the optimized workload for a particular GCF memory configuration. On completion of a \texttt{k6} load test for a particular function, the data collector component of \textit{Optimus} queries the GCF monitoring metrics for the function and writes them to a CSV file as described in \S\ref{sec:dataacq}. The data is sampled at a granularity of 10s supported by GCF. For a particular function and GCF memory configuration, the average execution time is obtained by calculating the weighted average of the number of function invocations and the mean execution time of the function (see Table~\ref{tab:metrics}). To compare memory consumption, we use the default GCF monitoring metric, i.e., Memory usage and average it across all the available datapoints. The  obtained speedup and average memory usage for the different workloads for the different available GCF memory configurations is shown in Figure~\ref{fig:perfmemcomp}. We report all performance results for double precision floating point operations. 

\thispagestyle{empty}




\begin{figure*}[t]
\centering
    \begin{subfigure}{0.3\textwidth}
    \centering
        \includegraphics[width=\columnwidth]{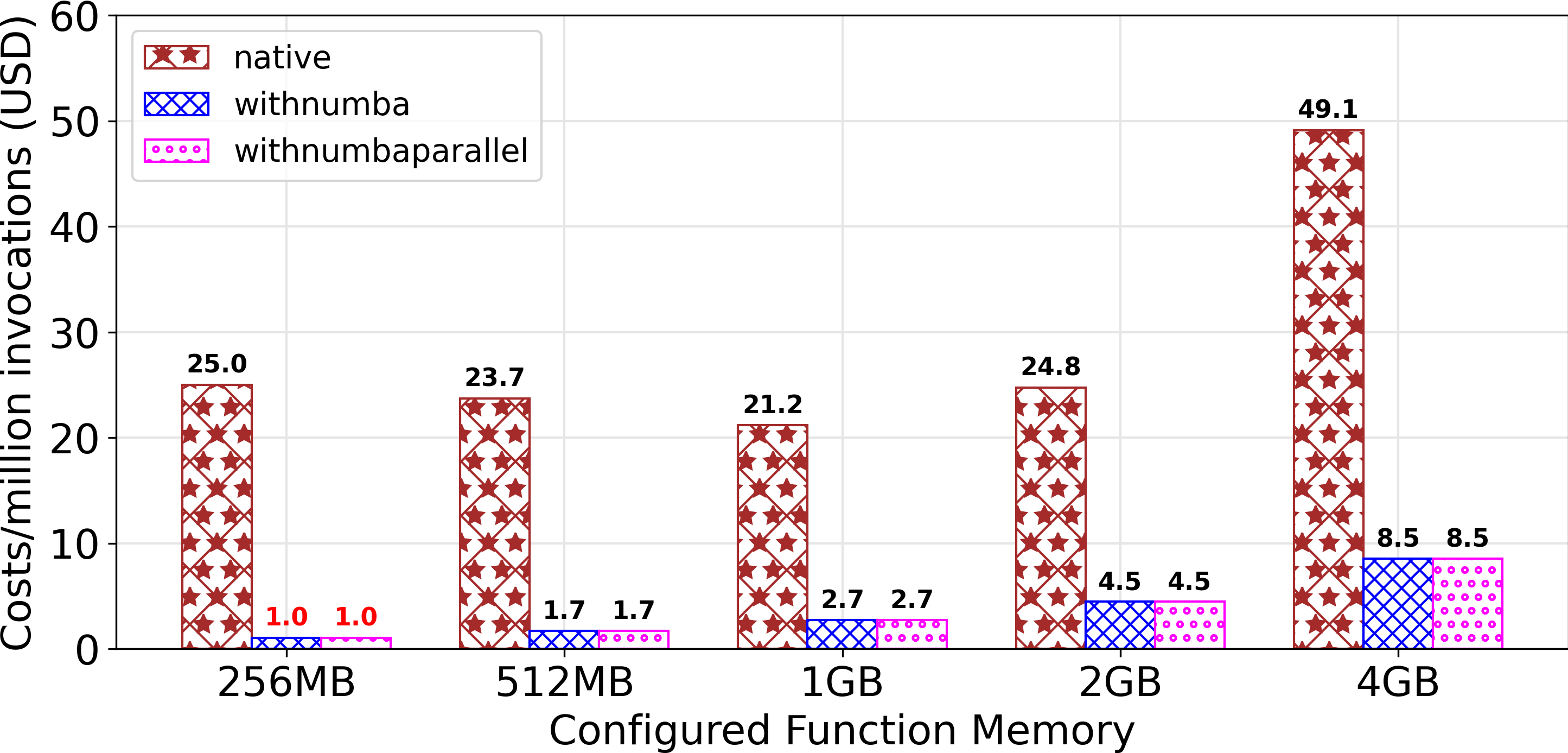}
        \caption{Floatbenchmark.}
        \label{cost:float}
    \end{subfigure}
    \begin{subfigure}{0.3\textwidth}
    \centering
        \includegraphics[width=\columnwidth]{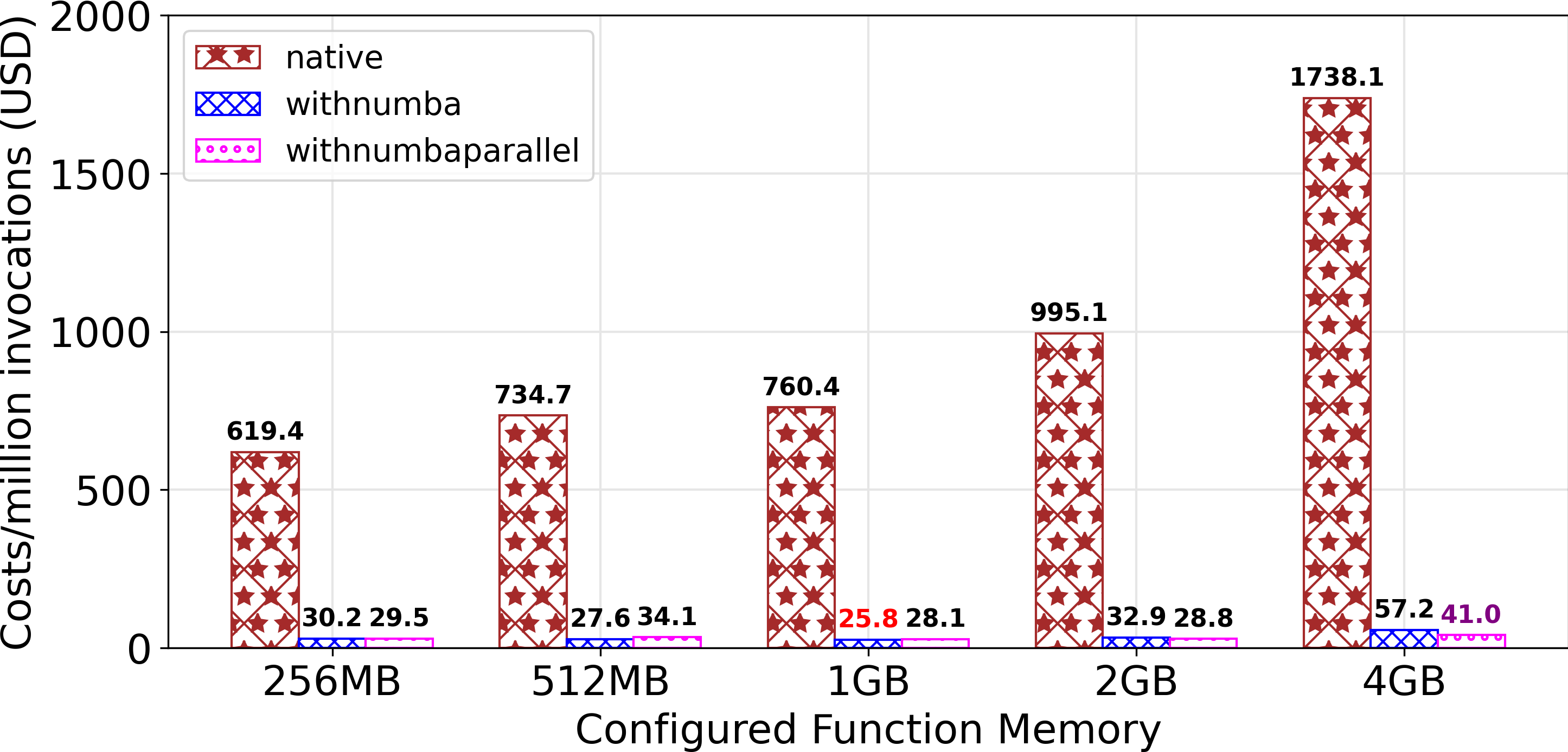}
        \caption{Mcbenchmark.}
        \label{cost:mcb}    
    \end{subfigure}
    \begin{subfigure}{0.3\textwidth}
    \centering
        \includegraphics[width=\columnwidth]{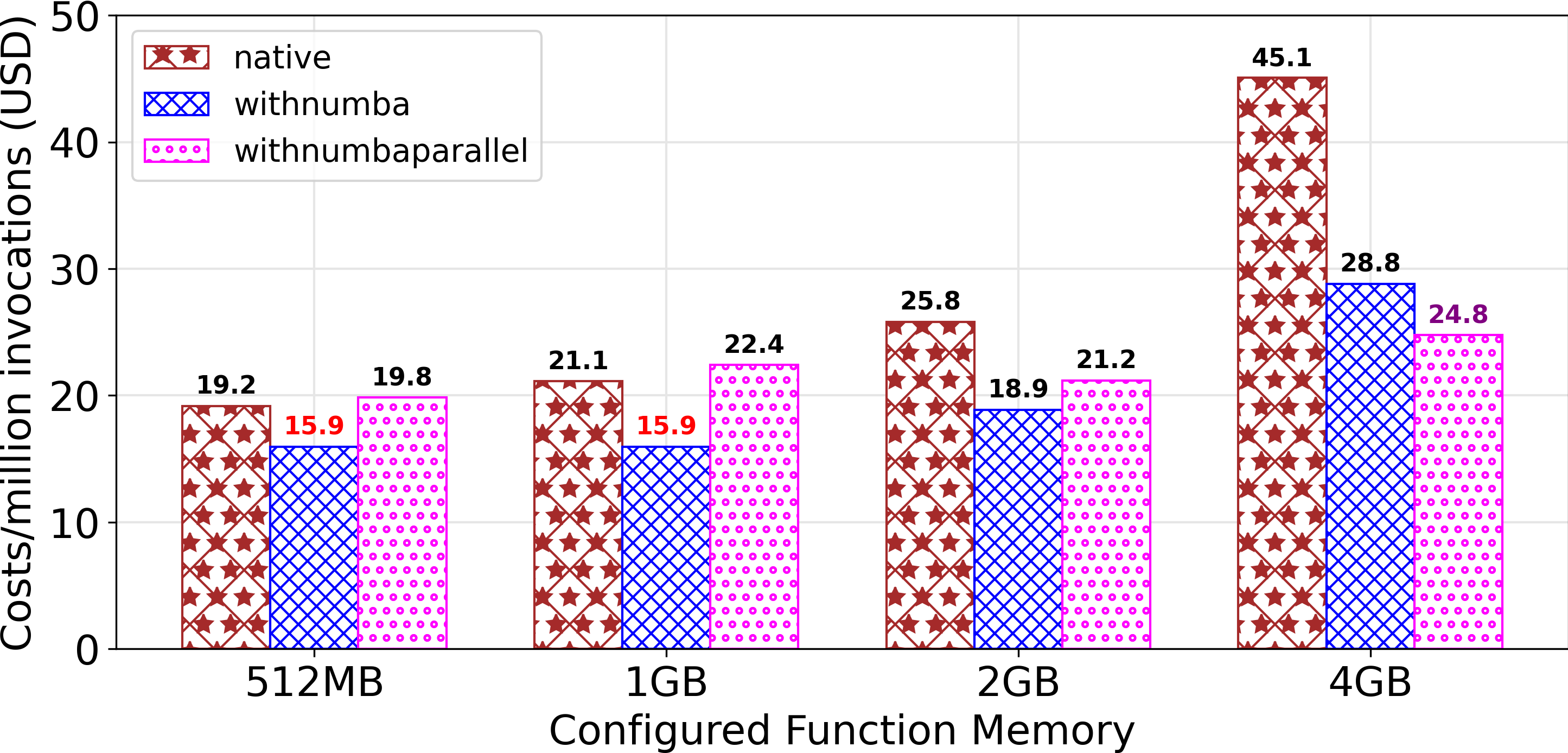}
        \caption{Imageprocessing}
        \label{cost:imgp}    
    \end{subfigure}
    \begin{subfigure}{0.3\textwidth}
    \centering
        \includegraphics[width=\columnwidth]{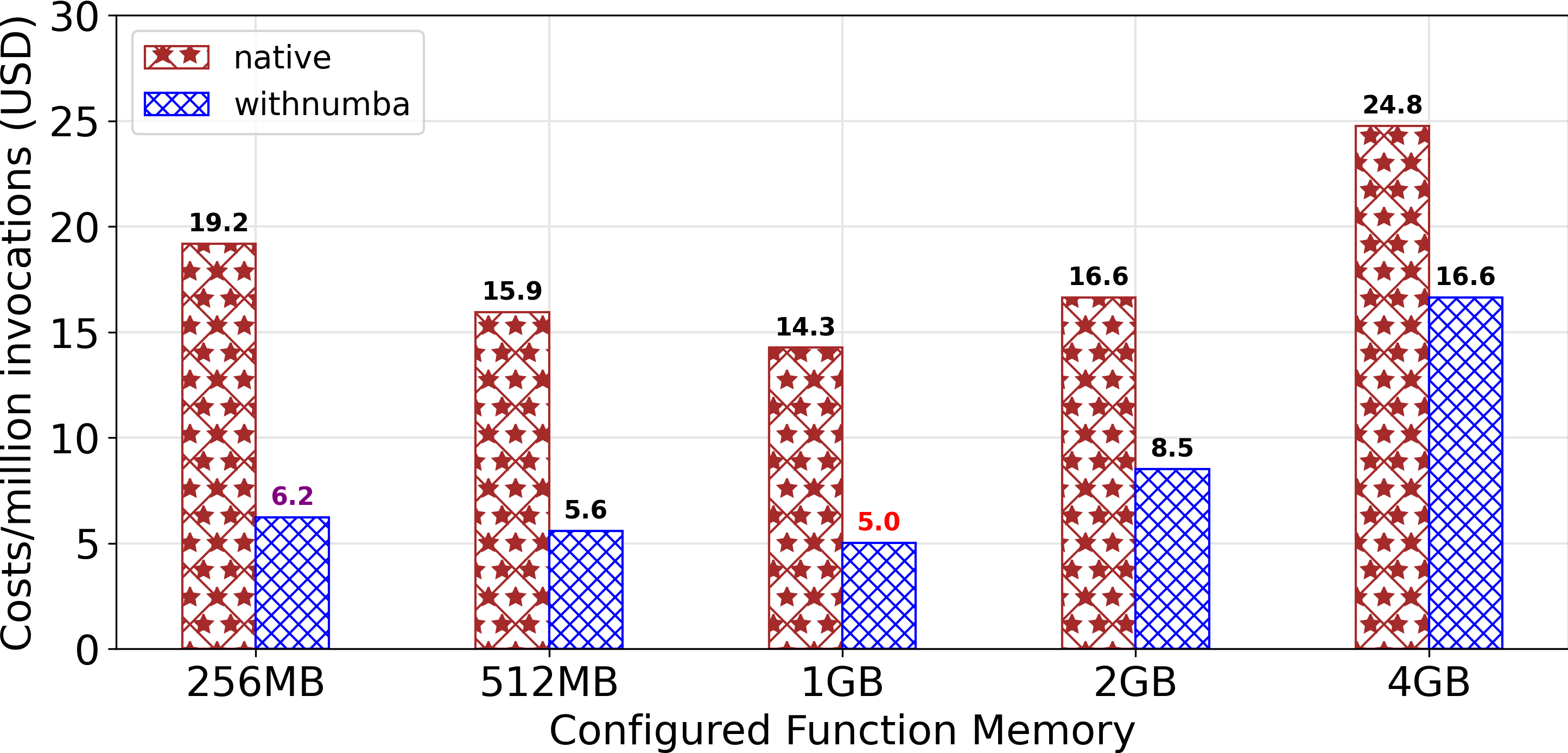}
        \caption{Logistic Regression}
        \label{cost:lr}    
    \end{subfigure}
    \begin{subfigure}{0.3\textwidth}
    \centering
        \includegraphics[width=\columnwidth]{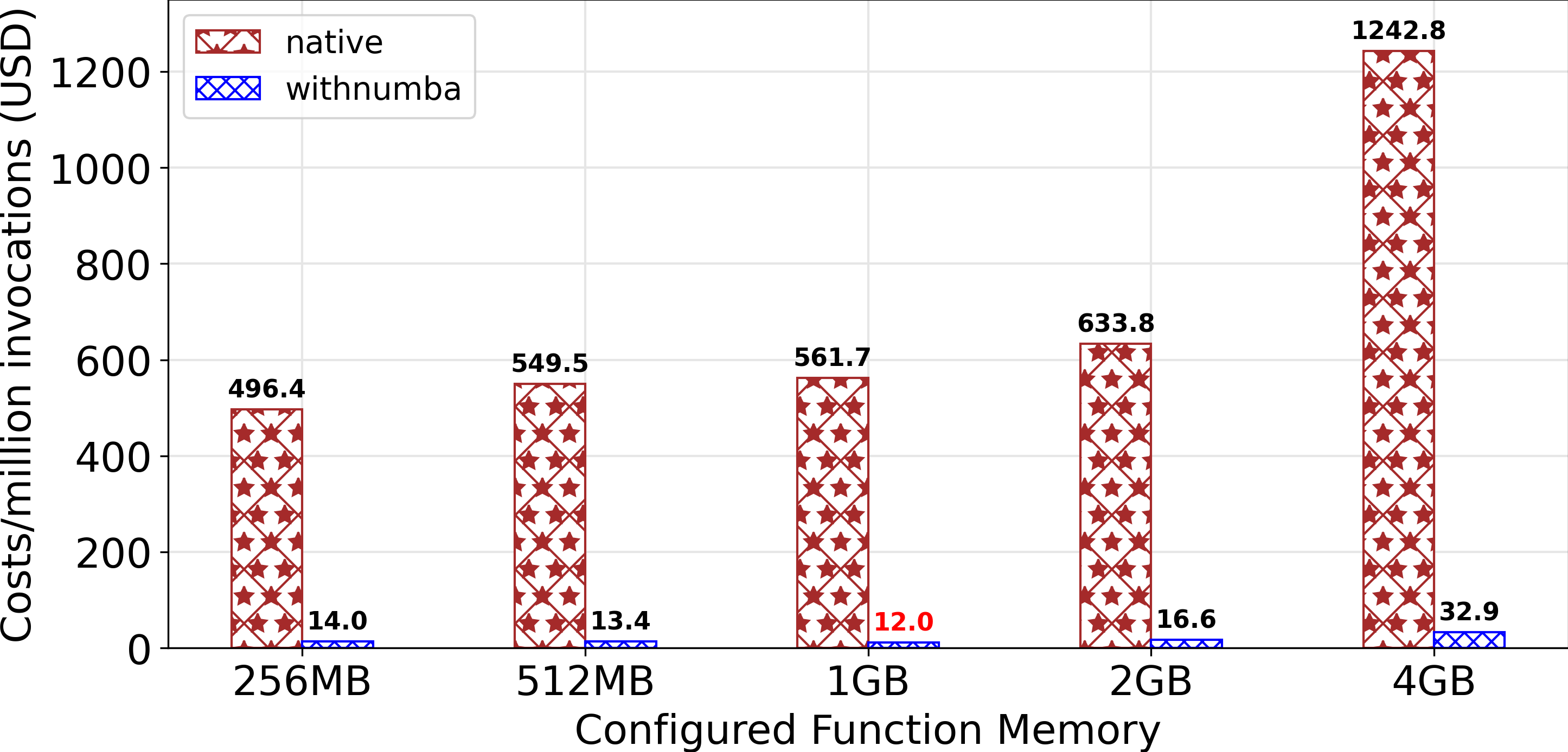}
        \caption{Nbody}
        \label{cost:nb}    
    \end{subfigure}
    \begin{subfigure}{0.3\textwidth}
    \centering
        \includegraphics[width=\columnwidth]{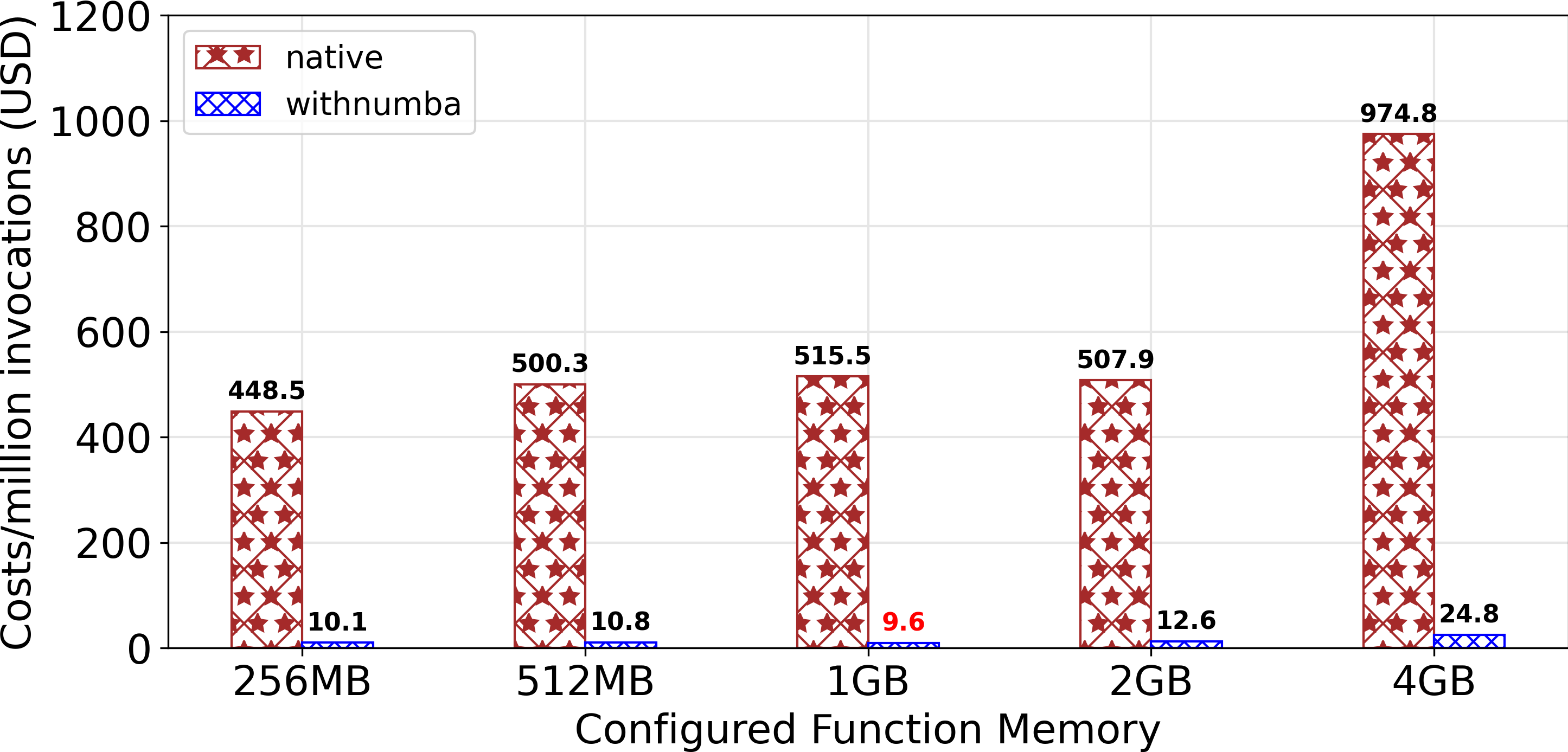}
        \caption{Kerneldensityestimate}
        \label{cost:kde}    
    \end{subfigure}
    
    \caption{Comparison of cost per million function invocations (in USD) of the six FaaS workloads as compared to their native implementations for the different memory configurations on GCF. The cost values highlighted with red represent the minimum values obtained across the different memory configurations, while the cost values highlighted with purple (if present and different) represent the values wrt the maximum percentage cost savings.}
    \label{fig:cost_compare}
\end{figure*}

\thispagestyle{empty}
For the \textit{Floatbenchmark}, we obtained a geometric mean speedup of $107$x, $113$x across the different memory configurations for the single-threaded and parallel versions optimized with Numba respectively. The maximum speedup for both versions, i.e., $311$x, $247$x is obtained for the memory configuration of $256$MB as shown in Figure~\ref{perf_float}. The main reason for the significant increase in the performance of the FaaS functions optimized with Numba is the generation and execution of machine code as described in \S\ref{sec:background}. On the other hand, for the native FaaS function, Python automatically generates bytecode which is executed by the default bytecode interpreter~\cite{sanner1999python}. For a given code statement, the generated bytecode contains substantially more CPU instructions as compared to the generated machine code by LLVM leading to a degradation in performance. As shown in Figure~\ref{perf_float}, the obtained speedup for both the optimized versions decreases when more memory is configured. This is because with increasing memory configuration GCF increases the number of CPU cycles allocated to a function~\cite{princinggcf}. As a result, the performance of the native FaaS function is enhanced. For the \textit{Floatbenchmark}, the optimized functions do not benefit from an increase in the number of CPU cycles since the generated vectorized code, due to auto-vectorization by LLVM, is more limited by memory bandwidth than the scalar native code. Although the underlying provisioned VMs are configured with two vCPUs (\S\ref{sec:processor_arch}), we do not observe an increase in speedup for the parallel function as compared to the single-threaded function for all memory configurations. This is because GCF uses a process-based model for resource management, where each function has a fixed memory and allocated CPU cycles. Since Intel-TBB follows a \textit{fork-join} model for parallel execution, the generated threads are inherently limited by the resource constraints of the parent process. We observe that the speedup of the parallelized function as compared to the single-threaded version increases with the increase in the allocated CPU clock cycles.

\thispagestyle{empty}

We obtained a geometric mean speedup of $28$x, $31$x for the single-threaded and parallelized versions of the \textit{Mcbenchmark} across the different memory configurations as shown in Figure~\ref{perf_mcb}. In contrast to Figure~\ref{perf_float}, we observe a different trend for the obtained speedup values due to memory bandwidth not being a bottleneck. The obtained speedup for the single-threaded function remains almost the same, i.e., $30$x when the function is configured with a memory of $1GB$ and higher. On the other hand, the speedup obtained for the parallelized function increases with increasing memory configuration, with the maximum obtained value of $52$x with $4GB$ of memory. For the \textit{Image Processing} workload, we obtained an average speedup of $1.39$x, $1.19$x across the different memory configurations for the single-threaded and parallelized versions respectively. The speedup values obtained are comparatively small since the native implementation of the benchmark uses the Python \texttt{Pillow} library (\S\ref{sec:benchmarks}). The \texttt{Pillow} library is implemented in C and can be directly called from the native Python interpreter~\cite{pycref}. As shown in Figure~\ref{perf_imgp}, the single-threaded Numba optimized \textit{Image processing} function performs better than the native implementation due to LLVM compiler optimizations, and vectorization using the highest underlying SIMD instruction set (\S\ref{sec:perf_tips}). In contrast, \texttt{Pillow} is pre-compiled and generic to \texttt{x86-64}. This means that the vector instructions generated will be for the Streaming SIMD Extensions (SSE) instruction set, which assumes a 128 bit SIMD unit length (\S\ref{sec:perf_aspects}). The parallelized Numba optimized function performs worse than the native implementation for the memory configurations  $512$MB, $1$GB, due to limited CPU clock cycles and parallelization overhead. Similar to Figure~\ref{perf_mcb}, the performance of the parallelized function improves with a higher memory configuration.

We observe a geometric mean speedup of $2.78$x across the different memory configurations for the \textit{Logistic Regression} (LR) function optimized with Numba. The maximum speedup value of $3.23$x is obtained for the memory configuration of $256$MB as shown in Figure~\ref{perf_lr}. The native implementation of the LR function uses \texttt{Numpy} which is pre-compiled for \texttt{x86-64}. As a result, the Numba optimized function outperforms the native implementation. For the optimized \textit{Nbody} and \textit{Kernel Density Estimate} functions we observe a geometric mean speedup of $46$x, $53$x across the different GCF memory configurations respectively. We observe a maximum speedup of $51$x, $61$x for the optimized \textit{Nbody} and KDE functions for the memory configurations of $2$GB, $1$GB as shown in Figures~\ref{perf_nb} and~\ref{perf_kde}.

For all benchmarks, we observe that the average memory usage of the Numba optimized functions is higher than their native implementations as shown in Figures~\ref{perf_float},~\ref{perf_mcb},~\ref{perf_imgp},~\ref{perf_lr},~\ref{perf_nb}, and~\ref{perf_kde}. This can be attributed to (i) additional variables required for Numba's internal compilation workflow (\S\ref{sec:background}), (ii) additional module dependencies such as LLVM, \texttt{icc\_rt}, and (iii) in-memory caching of the generated machine code. The memory required for the Numba parallelized functions is more as compared to the single-threaded functions because of the additional \texttt{intel-tbb} library. Note that, due to the presence of coarse grained memory profiles and billing policy adopted by GCF~\cite{princinggcf}, users will be charged based on the configured memory, irrespective of the function memory usage. The memory consumption of 
of the different functions is similar across the different memory configurations leading to memory over-provisioning.

\begin{figure}[t]
\centering
\includegraphics[width=0.7\columnwidth]{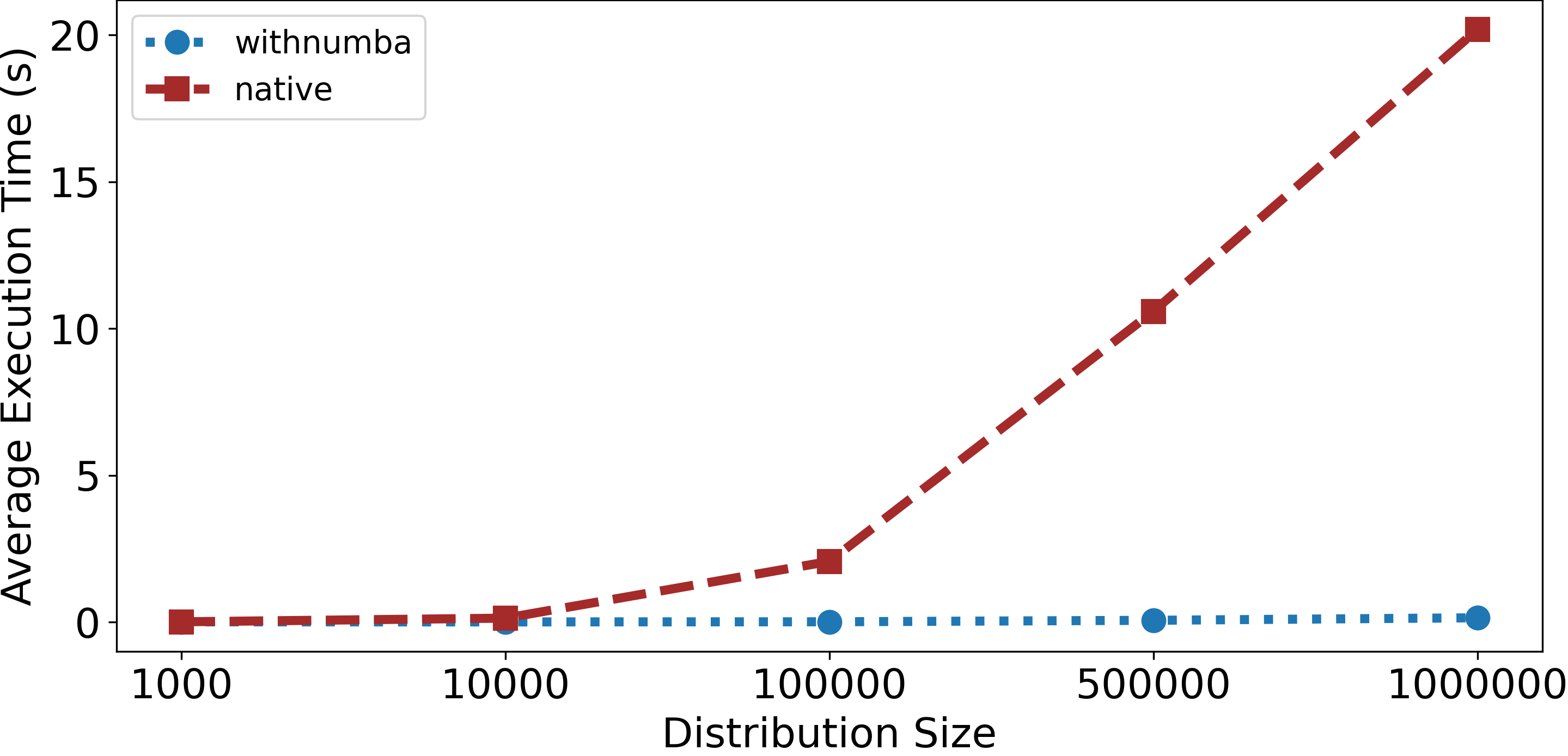}
\caption{Comparison of the effect of increasing the \textit{distribution size} on the average execution time for the optimized and native versions of the \textit{Kde} FaaS workload when deployed with $256$MB on the \texttt{us-west2} region.}
\label{fig:scale}
\end{figure}

Another advantage of the JIT compilation by LLVM supported by Numba is the explicit avoidance of creation of temporary arrays. Figure~\ref{fig:scale} shows the effect of increasing the argument, \textit{distribution size} on the performance of the \textit{KDE} workload. The native implementation of the \textit{KDE} function is done using Numpy as described in \S\ref{sec:benchmarks}. For small distribution sizes, the native implementation performs similar to the Numba optimized function. However, with increasing distribution size we observe an exponential increase in the average execution time.  This can be attributed to the repeated allocation, deallocation of temporary internal Numpy arrays~\cite{numpyint}, which are avoided by Numba.

\subsection{Comparing costs}
\label{sec:setup}

\thispagestyle{empty}

Figure~\ref{fig:cost_compare} shows the cost per million invocations of the optimized FaaS workloads as compared to their native implementations for the different memory profiles on GCF. To compute the invocation cost of a particular function and GCF memory configuration, we use the obtained average execution time (\S\ref{sec:perfmem}) and round it up to the nearest $100$ms increment.
Following this, we use the rounded average execution time to calculate the function compute time in terms of the units GB-Second and GHz-Second. The compute time depends on the configured memory and the allocated CPU clock cycles (defined by GCF). For instance, with a memory configuration of $256$MB, the associated clock cycles is $400$MHz~\cite{princinggcf}. GCF defines a fixed price for one second of compute time depending on the region where the function is deployed. We use the pre-defined price values for calculating the function compute cost. In our calculation, we exclude the cost for free tiers and networking. As a result, a fixed price of \$$0.40$ per million invocations is added to the calculated function compute cost.

For the \textit{Floatbenchmark}, we observe $88$\% average cost savings for the single-threaded and parallelized functions across the different memory configurations. Although there is a difference in the obtained speedup for the two different optimized functions (see Figure~\ref{perf_float}), the cost values obtained are the same as shown in Figure~\ref{cost:float}. This can be attributed to the coarse-grained 100ms billing intervals used by GCF. Note that, for FaaS providers such as AWS Lambda and Azure functions with 1ms billing intervals the costs obtained for the parallelized version will be less when configured with memory greater than 256MB. The minimum cost and maximum cost savings of \$$1.0$ and $95.8$\% are obtained for the memory configuration of $256$MB corresponding to the maximum obtained speedup for the two functions. We observe $96.2$\%, $96.4$\% average cost savings for the two Numba optimized functions of the \textit{Mcbenchmark}. The minimum cost value of \$$25.8$ is obtained for the single threaded function when configured with $1$GB of memory as shown in Figure~\ref{cost:mcb}. The maximum cost savings of $97.64$\% is obtained with a memory configuration of $4$GB for the parallelized function.


We observe $26.1$\% average cost savings for the single-threaded \textit{Image processing} function across the different memory configurations. The cost values obtained for the parallelized function are higher as compared to the native implementation for the memory configurations $512$MB and $1$GB respectively. But, they decrease when higher memory is configured as shown in Figure~\ref{cost:imgp}. The minimum cost value of \$$15.9$ is obtained for the single-threaded function when configured with either $512$MB, or $1$GB of memory. The maximum cost savings of $45$\% is obtained for the parallelized function when configured with $4$GB of memory. For the \textit{Logistic Regression} workload, we observe $55.8$\% average cost savings for the Numba optimized function across the different memory configurations. The minimum cost value of \$$5.0$ is obtained for the memory configuration of $1$GB, while the maximum cost savings of $67.6$\% is obtained for the memory configuration of $256$MB. For the optimized \textit{Nbody} function, we observe $97.47$\% average cost savings across the different memory configurations. The minimum cost and maximum cost savings of \$$12.0$ and $97.8$\% are obtained for the memory configuration of $1$GB as shown in Figure~\ref{cost:nb}. We observe $97.75$\% average cost savings for the optimized \textit{KDE} function across the different memory configurations. Similar to the optimized \textit{Nbody} function, the minimum cost value and maximum cost savings of \$$9.6$ and $98.1$\% are obtained for the memory configuration of $1$GB as shown in Figure~\ref{cost:kde}. 

\thispagestyle{empty}
Although the speedup obtained for the different optimized function varies across the different memory configurations (\S\ref{sec:perfmem}), we do not observe a significant difference in costs for the Numba optimized functions across the memory configurations as shown in Figure~\ref{fig:cost_compare}. GCF offers the possibility of unlimited scaling of function instances to meet user demand~\cite{gcfscaling}. To avoid memory over-provisioning and due to the significant speedup obtained with Numba for the lowest possible memory configuration for a particular function, the minimum memory configuration can always be selected. Moreover, we observe that parallelization of functions is only beneficial when configured with a memory of $2GB$ and higher because of constraints on the allocated CPU clock cycles.






\subsection{Effect of heterogeneity in the underlying processor architectures on performance}
\label{sec:setup}
\begin{figure}[t]
\centering
\begin{subfigure}{0.49\columnwidth}
    \centering
        \includegraphics[width=\columnwidth]{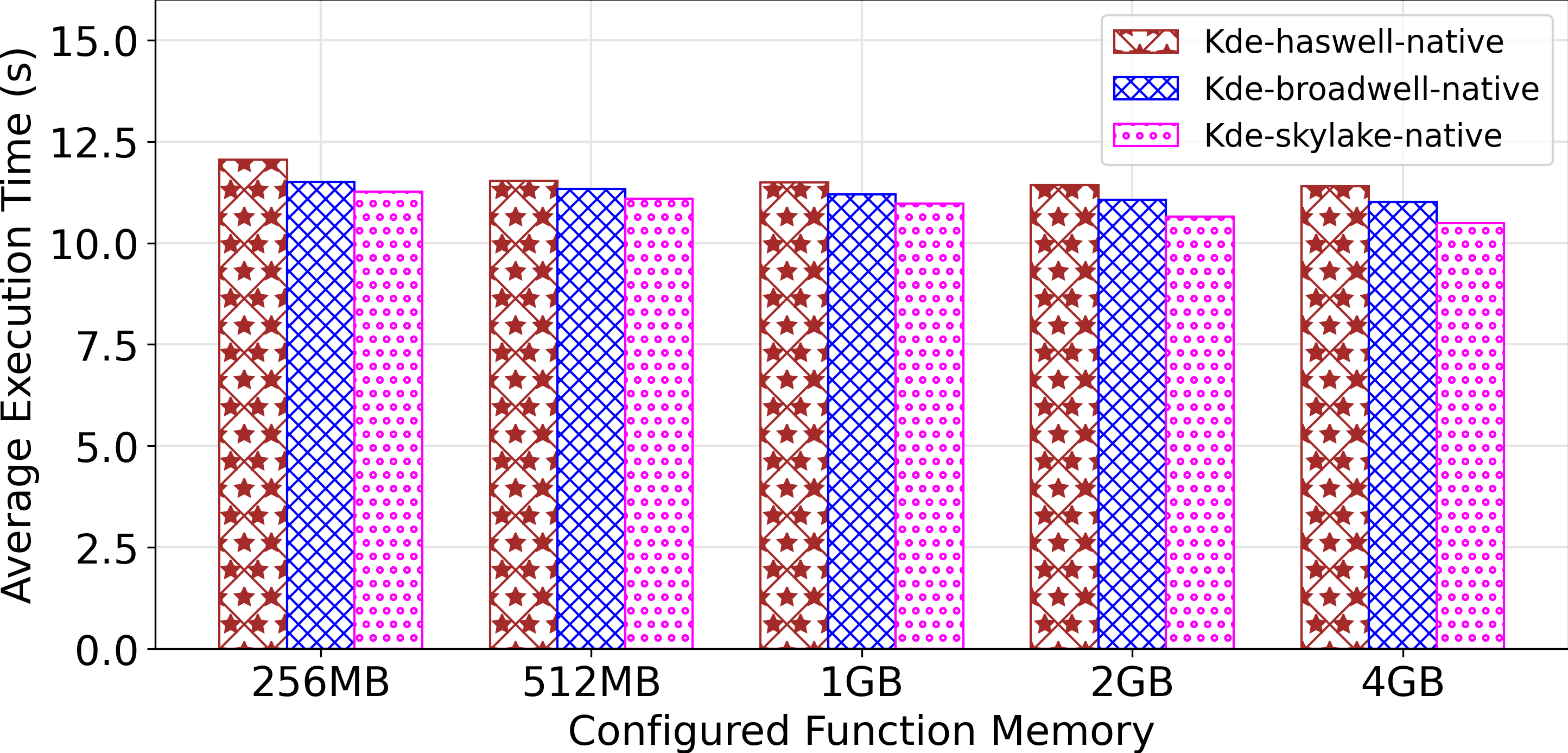}
        \caption{Kde Native.}
        \label{fig:native}    
\end{subfigure}
\begin{subfigure}{0.49\columnwidth}
    \centering
        \includegraphics[width=\columnwidth]{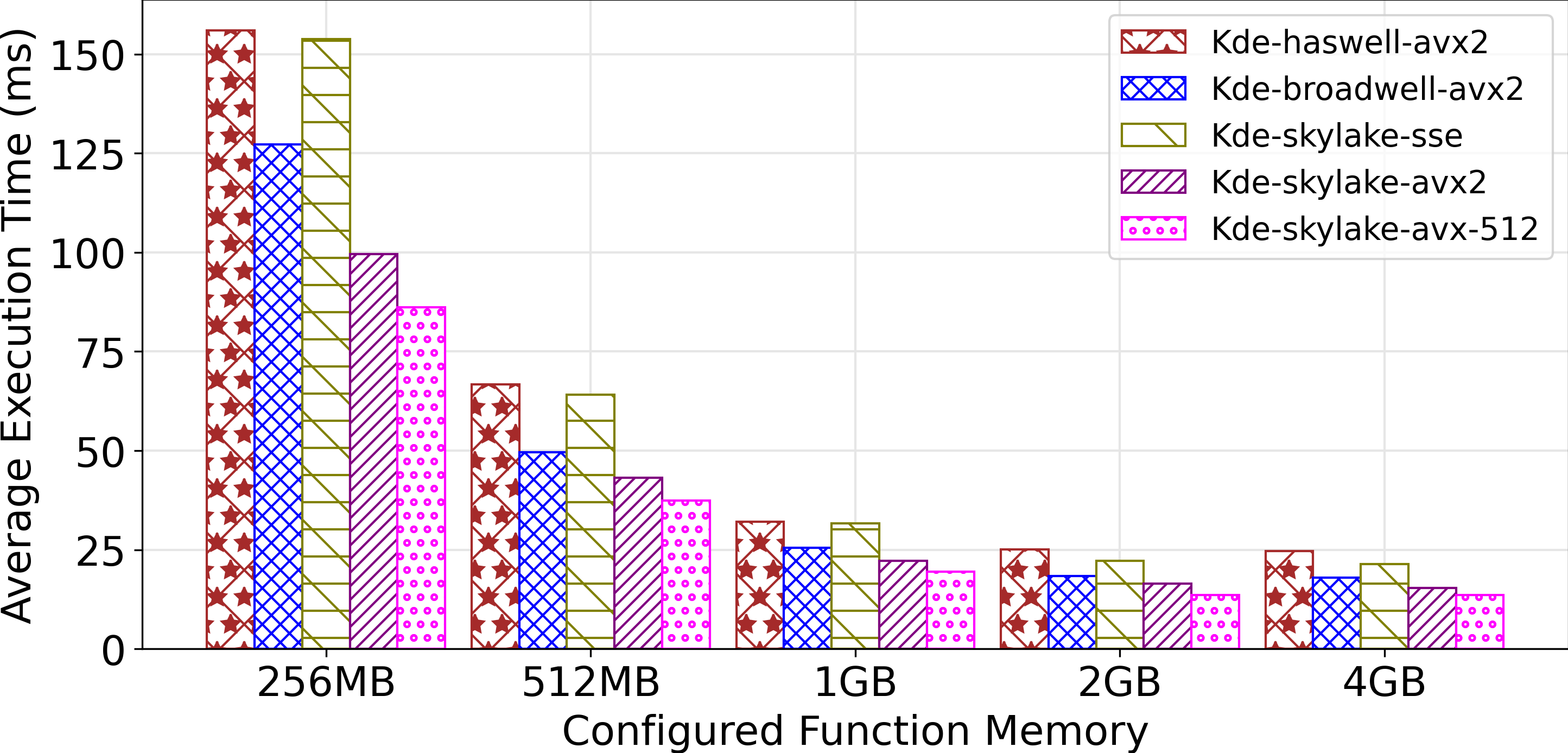}
        \caption{Kde Numba}
        \label{fig:optimized}
\end{subfigure}

\caption{Comparison of the execution times for the optimized and native versions of the \textit{Kde} FaaS workload for the different underlying processor architectures. The functions were deployed on the \texttt{asia-northeast1} region.}
\label{fig:heterocompare}
\end{figure}

To analyze the effect of different processor architectures on the performance of a FaaS function, we use the \textit{Kernel Density Estimate} (KDE) workload and deploy it for all supported memory configurations in the \texttt{asia-northeast1} region. We chose this region since it had the greatest heterogeneity and prevalence of the three processor architectures (\S\ref{sec:processor_arch}). We instrumented the KDE workload to compute the execution time required for calculating the estimate at the evaluation point (\S\ref{sec:benchmarks}) given as input. The processor architecture is determined similarly as described in \S\ref{sec:processor_arch}. The different attributes are collated and returned as a JSON response. As described in \S\ref{sec:perf_aspects}, Numba automatically generates SIMD instructions for highest underlying instruction set. However, to emphasize the importance of generating architecture-specific code, we modified the Numba configuration to generate only AVX-2 and SSE instructions on the Skylake processor. Figure~\ref{fig:optimized} shows the average execution time for the different processor architectures and SIMD instruction sets across the different memory configurations for the Numba optimized \textit{KDE} function.


For all processor architectures the average execution time decreases with increasing memory configuration since more compute is assigned. For the native \textit{KDE} implementation (see Figure~\ref{fig:native}), the Skylake processor obtains a speedup of $1.10$x, $1.03$x, on average across all memory configurations as compared to the Haswell and Broadwell processors. On the other hand, for the Numba optimized function, we observe an average speedup of $1.79$x, $1.36$x for the Skylake processor (with AVX-512) as compared to the Haswell and Broadwell processors respectively. Although, the native \textit{KDE} function implementation uses \texttt{Numpy} which is pre-compiled for \texttt{x86-64}, i.e., the generated vector instructions will use the SSE SIMD instruction set (\S\ref{sec:perfmem}), we observe a difference in performance for the different architectures. This is because of several  microarchitectural improvements to the Skylake processor~\cite{schone2019energy}. The difference in performance is more significant for the Numba optimized function because the LLVM compiler in Numba autovectorizes the jitted function in the \textit{KDE} workload to generate instructions using the AVX-512 instruction set on the Skylake processor and using the AVX-2 instruction set on the Haswell and Broadwell processors. As a sanity check, we also confirmed this by examining the assembly code of the jitted function and checking the registers used in the generated vector instructions (\S\ref{sec:perf_aspects}). The Broadwell processor obtains a speedup of $1.03$x, $1.31$x on average across all memory configurations as compared to the Haswell processor for the native and Numba optimized functions respectively. This can be attributed to a higher Instructions per cycle (IPC) value and reduced latency for floating point operations as compared to the Haswell processor~\cite{broadwell}.

In comparison to the Numba optimized function with SSE and AVX-2 generated instructions on the Skylake processor, the version with AVX-512 instructions obtains a best speedup of $1.67$x and $1.16$x on average across all memory configurations respectively. Moreover, the SSE version on the Skylake processor is $1.23$x slower on average than the optimized version with AVX-2 instructions on the Broadwell processor. Although there is an illusion of homogeneity in most public FaaS offerings, the actual performance of a FaaS function can vary depending on the underlying architecture of the provisioned VM where the function instance is launched. As a result, the cost incurred for the same function will also vary.

\thispagestyle{empty}
\section{Conclusion \& Future Work}
\label{sec:conclusion}
In this paper, we adapted and optimized a representative set of six compute-intensive FaaS workloads with Numba, i.e., a JIT compiler based on LLVM. We determined the different processor architectures used by GCF namely Haswell, Broadwell, and Skylake in the underlying provisioned VMs on which the function instances are launched. Furthermore, we identified the prevalence of these architectures across the 19 available GCF regions. Moreover, we demonstrated the use of underlying VM configuration, i.e., number of vCPUs for parallelizing FaaS functions. We deployed the optimized workloads on GCF and presented results wrt performance, memory consumption, and costs. We showed that optimizing FaaS functions with Numba can improve performance by $44.2$x and save costs by $76.8$\% on average across the six functions. We investigated the effect of the underlying heterogeneous processor architectures on the performance of FaaS functions. We found that the performance of a particular optimized FaaS function can vary by $1.79$x, $1.36$x on average depending on the underlying processor. Moreover, under-optimization of a function based on the underlying architecture can degrade the performance by a value of $1.67$x. In the future, we plan to investigate strategies for caching the compiled optimized machine code to reduce the startup times of functions.



\section{Acknowledgement}
This work was supported by the funding of the German Federal Ministry of Education and Research (BMBF) in the scope of the Software Campus program. Google Cloud credits were provided by the Google Cloud Platform research credits.

\bibliographystyle{IEEEtran}
\thispagestyle{empty}
\bibliography{parallelpgm}


\end{document}